\documentclass[fleqn,usenatbib]{mnras}

\usepackage{newtxtext,newtxmath}

\usepackage[T1]{fontenc}

\DeclareRobustCommand{\VAN}[3]{#2}
\let\VANthebibliography\thebibliography
\def\thebibliography{\DeclareRobustCommand{\VAN}[3]{##3}\VANthebibliography}

\usepackage{graphicx}	
\usepackage{amsmath}	
\usepackage{enumitem}
\usepackage{comment}
\usepackage{adjustbox}

\let\oldAA\AA
\renewcommand{\AA}{\text{\normalfont\oldAA}}

\title[Sputtering Process On Icy grains]{The Cosmic-Ray Induced Sputtering Process On Icy Grains}

\author[\"{O}zg\"{u}n Arslan et al.]{
\"{O}zg\"{u}n Arslan,$^{1,2}$\thanks{E-mail: oarslan@erciyes.edu.tr}
Seyit Hocuk,$^{3}$
Paola Caselli$^{4}$
and \.{I}brahim K\"{u}\c{c}\"{u}k$^{1,2}$
\\
$^{1}$ Department of Astronomy and Space Sciences, Science Faculty, Erciyes University, 38030, Melikgazi, Kayseri, Türkiye\\
$^{2}$ Astronomy and Space Sciences Observatory and Research Center (Uzaybimer), Erciyes University, 38281, Talas, Kayseri, Türkiye\\
$^{3}$ CentERdata, Institute for data collection and research, Professor de Moorplein 524-525, 5037 DR Tilburg, The Netherlands\\
$^{4}$ Center for Astrochemical Studies, Max-Planck-Institute for Extraterrestrial Physics, Gießenbachstr. 1, D-85741 Garching, Germany\\}

\date{Accepted XXX. Received YYY; in original form ZZZ}

\pubyear{2022}

\begin{document}
\label{firstpage}
\pagerange{\pageref{firstpage}--\pageref{lastpage}}
\maketitle

\begin{abstract}
In molecular cloud cores, the cosmic ray (CR) induced sputtering via CR ion-icy grain collision is one of the desorption processes for ice molecules from mantles around dust grains. The efficiency of this process depends on the incident CR ion properties as well as the physicochemical character of the ice mantle.
Our main objective is the examination of the sputtering efficiency for H$_2$O and CO ices found in molecular cloud cores.
In the calculation routine, we consider a multi-dimensional parameter space that consists of thirty CR ion types, five different CR ion energy flux distributions, two separate ice mantle components (pure H$_2$O and CO), three ice formation states, and two sputtering regimes (linear and quadratic). 
We find that the sputtering behavior of H$_2$O and CO ices is dominated by the quadratic regime rather than the linear regime, especially for CO sputtering. The sputtering rate coefficients for H$_2$O and CO ices show distinct variations with respect to the adopted CR ion energy flux as well as the grain size-dependent mantle depth. The maximum radius of the cylindrical latent region is quite sensitive to the effective electronic stopping power. The track radii for CO ice are much bigger than H$_2$O ice values. In contrast to the H$_2$O mantle, even relatively light CR ions ($Z \geq 4$) may lead to a track formation within the CO mantle, depending on ${\rm S_{\rm e,eff}}$. 
We suggest that the latent track formation threshold can be assumed as a separator between the linear and the quadratic regimes for sputtering. 
\end{abstract}

\begin{keywords}
astrochemistry -- ISM: molecules -- ISM: clouds -- (ISM:) cosmic rays
\end{keywords}



\section{Introduction}
 Atomic and molecular species in the gas-phase found in dense interstellar environments such as molecular cloud cores can actively condense on carbonaceous or silicate grains \citep{walmsley2004complete, steinacker2015grain,noble2017two}.  If grain sizes are large enough ($\geq 0.03$ $\mu$m) to prevent stochastic impulse heating induced by energetic agents, for instance, ultraviolet (UV) photons or cosmic ray (CR) ions \citep{herbst2005chemistry,draine2010physics,abplanalp2016exploiting}, the condensed species can turn into ice mantles on bare grain surfaces \citep{tielens2005physics,hollenbach2008water, oberg2011ices, hocuk2016chemistry}.

The ice mantles would normally stay on grain surfaces in cold environments since efficient thermal desorption processes are possible only for sufficiently high ambient medium temperatures \citep[$\ge$\,20\,K,][]{garrod2006formation,cuppen2017grain}. The gas-phase chemistry cannot simply explain  molecular species that have been observed in the gas phase in dense clump structures and molecular cloud cores \citep{vasyunin2013reactive, shingledecker2018cosmic,wakelam2021efficiency}.
Therefore, non-thermal desorption mechanisms are needed to explain the observed gas-phase abundances of some molecular species in these environments with typical conditions of $\rm T\leqslant 20\,K$, $\rm n_H \ge 10^{4} \rm ~cm^{-3}$, $A_{V} \ge 10 \rm ~mag$ \citep{reboussin2014grain,cazaux2016dust}.

For denser molecular cloud structures, such as the inner parts of dense clouds and starless cloud cores that are highly shielded from external sources with even the most intense UV radiation, one of the most significant desorption mechanisms of ice mantle molecules is through the heating of icy grains by CR particles. The desorption arises either by direct collisions \citep{de1973growth, jurac1998monte, bringa2007energetic, ivlev2015impulsive,kalvans2018temperature} or by indirect, CR-induced secondary UV irradiation \citep{shen2004cosmic,hollenbach2008water, caselli2012first}.  When a CR ion–icy grain collision occurs, the kinetic energy of the projectile CR ion can be partly deposited on the target grain via both elastic and inelastic interactions \citep {baragiola2003sputtering, sabin2009analytical}. The projectile CR ion energy loss is proportional to a quantity known as the stopping power \citep{sigmund1969theory}.  In general terms, the stopping power is the average energy loss of the charged projectile CR particle per path of the length of the target material \citep{meftah1993swift, johnson2013sputtering, shingledecker2018cosmic}. 

Depending on the characteristic properties of icy grains and CR ions, the grain heating processes induced by impinging CR ions that may lead to noticeable ice molecules desorption from grain surfaces have two different sub-regimes: whole-grain heating and hot spot heating \citep{leger1985desorption,hasegawa1993new,bringa2004new,dartois2015heavy,zhao2018effect}. 

Whole-grain heating is a consequence of several elastic and inelastic interactions between grains and CR ions and can be defined as the thermal diffusion process over the entire surface of the grain \citep{hasegawa1993new,zhao2018effect}. During the whole-grain heating process, the partially transferred energy from the incident CR ion to the target grain may cause a homogeneous and progressive rising of the grain surface temperature until the surface cooling driven by either the mantle evaporation or radiative emission can put a halt to it \citep{leger1985desorption,kalvans2015cosmic,kalvans2020evaporative,sipila2021revised}. 

In contrast to the whole-grain heating process, inelastic electronic interactions between target icy grain and incident CR ion may create a transiently and intensely heated local region on the ice mantle. These processes occur within very short picosecond timescales \citep{bringa2002sputtering,bringa2004new,mainitz2016irradiation,gupta2017role,anders2020ejection}. This type of CR-induced grain heating process is called hot spot heating \citep{leger1985desorption}. The locally and impulsively heated latent region from the hot spot heating occurs only around the radial trajectory of the incident CR ion passing through the target grain material \citep{dartois2015heavy,ivlev2015impulsive}. This local region is defined as the \textit{latent track} 
\citep{toulemonde1993thermal,lounis2008determination,szenes2011comparison,wesch2016ion}.

During hot spot heating (on timescales of $\mathrm{10}^{-17}$ to $ \mathrm{10}^{-9}\rm\,s$), the sequential and partial energy transfer interactions between electronic and atomic subsystems of the mantle within the latent track region may set a front motion for ice molecules, which results in the molecular ejection from the surface. This sublimation-like desorption process is known as \textit{electronic sputtering} \citep{sigmund1987mechanisms}. The hot spot-induced electronic sputtering plays a critical role in the efficient desorption of the non-polar and volatile ice. This sputtering also leads to the desorption of the polar and more-refractory ices with higher surface binding energies \citep{baragiola2003sputtering, mainitz2016irradiation,anders2019energetic,anders2019high,dartois2019non,anders2020ejection}.

The typical morphology of the latent track region across the direction of incident CR ion for an efficient molecular ejection is a continuous (or a single piece) cylinder with a radius that can vary a few ten $\mathring{\mathrm{A}}$ to hundreds of $\mathring{\mathrm{A}}$  \citep{beuve2003influence,toulemonde2004track,bringa2007energetic,lounis2008determination,
wesch2016ion,shingledecker2020cosmic}.  

There are many extensive studies on different types of icy grain-CR ion interactions, such as \citet{hasegawa1993new},  \citet{kalvans2019chemical}, \citet{kalvans2018temperature}, \citet{zhao2018effect} and \citet{sipila2020effect,sipila2021revised} for whole-grain heating and \citet{leger1985desorption}, \citet{schutte1991explosive}, \citet{shen2004cosmic} and \citet{ivlev2015impulsive} for explosive desorption of ice mantles. 

Several theoretical and experimental studies include the effects of incident CR ions on various target materials. These studies focus on the hotspot heating-induced sputtering process (linear or non-linear) and the details of latent track formation. Some of them are \citet []{erents1973desorption}, 
 \citet{brown1978sputtering,brown1980linear},  \citet{leger1985desorption},  \citet{schou1986erosion},  \citet{johnson1991linear},  \citet{bringa2002sputtering,bringa2004new}, 
 \citet {bringa2007energetic},    \citet{dartois2013swift,dartois2015heavy,dartois2018cosmic,dartois2019non,dartois2021cosmic},   \citet{anders2013impacts,anders2019energetic,anders2019high,anders2020ejection},  \citet{mainitz2016irradiation,mainitz2017impact}, \citet{shingledecker2020cosmic}, and \citet{silsbee2021ice}. 

However, all of these studies have certain limits due to the complex nature of the hot spot process as well as the selection criteria for calculations that are related to both the incident  CR ions  and  the structural properties of grain mantles. 

In this work, we investigate formation conditions of latent track regions with continuous cylindrical geometries on icy grains with different sizes during the hot spot heating for pure H$_2$O and CO ice mantles. To examine the efficiency of the CR-induced hot spot sputtering process on icy grains, according to the adopted environmental condition in a typical molecular cloud core, we calculate the sputtering yields and rate coefficients of H$_2$O and CO ice mantles for two sputtering regimes, namely the linear and the quadratic sputtering regimes. We expect that our study will contribute to the generation of more accurate and realistic chemical models for dense molecular cloud structures with the inclusion of the effects of hot spot heating-induced sputtering.

Our paper is divided into four sections. In Section 2, we describe the calculation steps needed to estimate the maximum latent ion track radii, the sputtering yields, and the sputtering rate coefficients. In Section 3, we share and discuss our results on the efficiency of the hot spot sputtering process. In Section 4, we summarise our results.

\section{METHODS \& MODELS}
\label{sec:MM} 
In this section, we describe the calculation routines needed to find the efficiency of the CR-induced hot spot sputtering process on icy grains. First, we describe the environmental conditions that we selected for the calculation routines.

\subsection{Environmental Conditions}
\label{sec:EC}
\begin{figure*}
\resizebox{0.96\textwidth}{!}{
\centering
\includegraphics[scale=0.48]{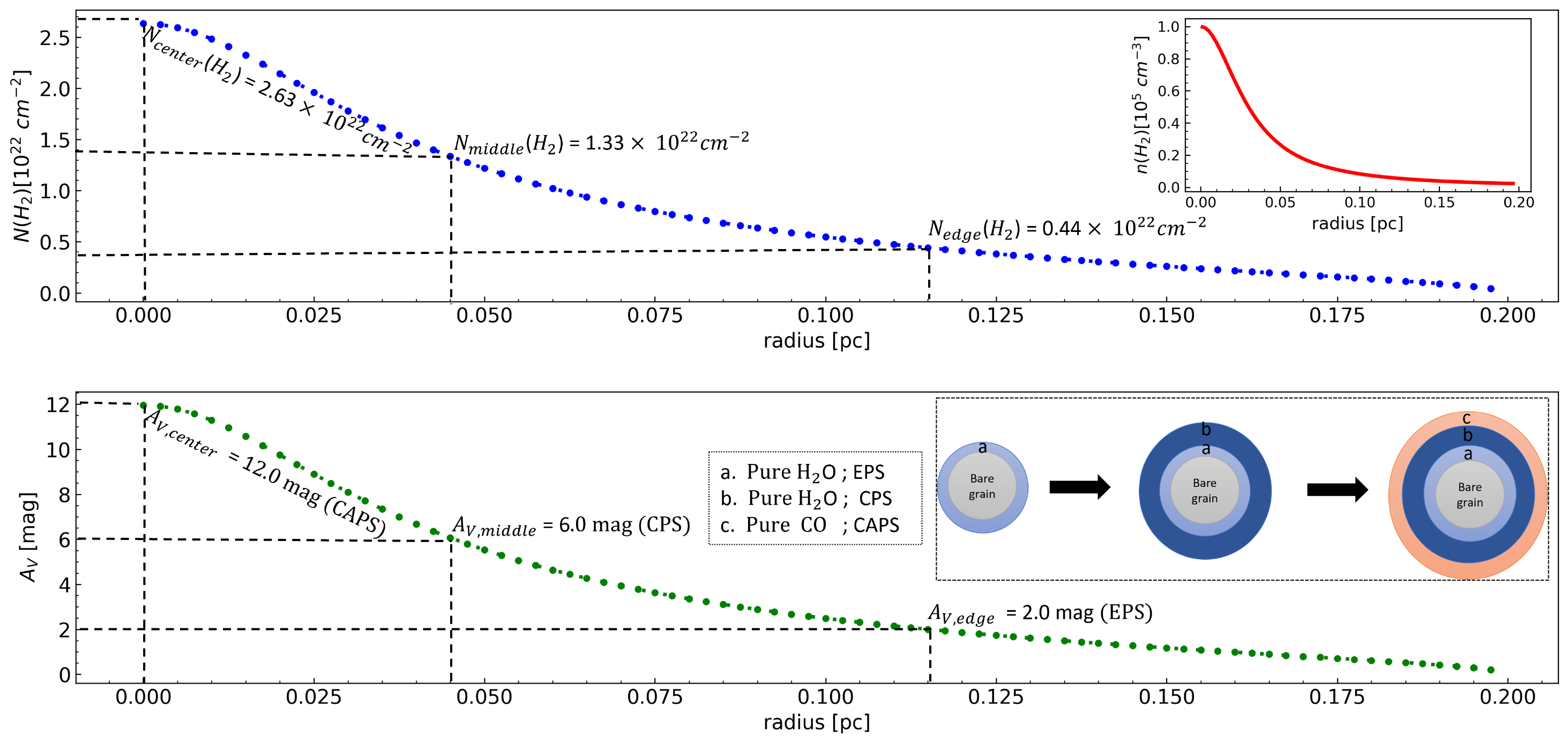}}
\caption{{\it{Top}}: The calculated molecular hydrogen density (solid red line in the inserted plot) and molecular hydrogen column density (blue dotted line) for a molecular cloud core that has a Plummer–like gas density distribution across the radial distance from the center to its edge.  To consider different ice formation states on surfaces of grain populations, we choose three conditions for the molecular cloud core based on three molecular hydrogen column densities defined as $N_{\rm edge}(\rm H_{2})$, $N_{\rm middle}(\rm H_{2})$, and $N_{\rm center}(\rm H_{2})$. 
{\it{Bottom}}: According to the adopted $N(\rm H_{2})/A_{V}$ ratio, the calculated visual extinction profile for cloud core, $A_{\rm V,edge}$ = 2.0 mag, $A_{\rm V,middle}$ = 6.0 mag and $A_{\rm V,center}$ = 12.0 mag, correspond to $N_{edge}(\rm H_{2})$ = $4.44 \times 10^{21}$ $\rm cm^{-2}$, $N_{\rm middle}(\rm H_{2})$ =  $1.33 \times 10^{21}$ $\rm cm^{-2}$ and $N_{\rm center}(\rm H_{2})$ = $2.63 \times 10^{22}$ $\rm cm^{-2}$, respectively.  $A_{V,middle}$  is the considered visual extinction value to separate H$_2$O ice formation between EPS and CPS. The illustrative drawing also shows three ice mantles with different thicknesses (a to c) that represent $A_{\rm V}$-dependent mantle evolution for different core conditions. The fixed ice mantle depth fractions are 11.37$\%$ (H$_2$O: EPS), 61.41$\%$ (H$_2$O: CPS), and 27.22$\%$  (CO: CAPS) for a, b, and c, respectively.}
\label{fig:fig1_NH_AV_relation}
\end{figure*}

We consider different parts of a typical molecular cloud core pertaining to the edge and center. Table~\ref{tab:table1_cond_core} summarizes the adopted environmental conditions. We define three ice formation states that are consistent for the edge and central regions. The states are: H$_2$O-dominated polar state in the edge (EPS), H$_2$O-dominated polar state in the center (CPS), and  CO-dominated apolar state in the center (CAPS). We illustrate EPS, CPS, and CAPS in Figure\,\ref{fig:fig1_NH_AV_relation}.

To determine the molecular hydrogen number density for the edge region, $n_{\rm edge}(\rm H_{2})$, we assume the cloud core exhibits a Plummer-like density distribution. This distribution has a characteristic flattening radius of $\rm R_{flat} = 0.03\,pc$, steepness index of $\rm p=2$ \citep{ysard2016mantle}, and a truncation radius of $\rm R_{out} = 0.2\,pc$ \citep{parikka2015physical}.

In evaluating the number density-dependent molecular hydrogen column density, $N(\rm H_{2})$, we use an analytical expression between $N(\rm H_{2})$ and $n(\rm H_{2})$ for the edge and the central regions, taking into account equation 27 of \citet{pineda2010relation}. For the conversion between the visual extinction $A_{V}$ and $N(\rm H_{2})$, we choose the ratio of $N(\rm H_{2})/A_{\rm V} = 2.2 \times 10^{21} \, \rm cm^{-2}\, \rm mag^{-1}$ \citep{guver2009relation}. This is slightly higher than some other commonly used values in the literature, that is $2 \times 10^{21} \, \rm cm^{-2}\, \rm mag^{-1}$ \citep[e.g.,][]{valencic2015interstellar,zhu2017gas}. 

Considering figure 9 of  \citet{hocuk2015interplay}, we adopt $A_{\rm V,edge}$ = 2.0 mag as the thin H$_2$O ice formation threshold at the EPS and $A_{\rm V,middle}$ = 6.0 mag as the thin to the thick ice formation threshold between the EPS and CPS. We use a linear correlation following figure 7 of \citet{boogert2015observations} to derive the H$_2$O ice column densities for the two states. 

We assume complete CO freeze-out occurs at $A_{\rm V,center}$ = 12.0 mag. According to this assumption, to obtain CO ice column density for CAPS, we employ the $A_{\rm V}$-dependent CO  column density approximation of \citet{pineda2010relation}, following their equation 22. 

In calculating the maximum mantle depths related to the different ice formation states, we use three $A_{V}$-dependent fixed mantle depth ratios, 0.114 (EPS), 0.614 (CPS), and 0.272 (CAPS) for each grain size population (see Section ~\ref{sec:Icy Grain} for details). 
\begin{table}
\caption{The adopted environmental parameters for edge and central regions of a typical molecular cloud core.}
\begin{tabular}{l |c |c |c }      
\hline\hline                 
 & Parameter &  Edge Region  &  Central Region \\
\hline
   1 & $n(\rm H_{2})  \; \rm cm^{-3}$    & $6.37 \times 10^3$      & $10^5$                   \\      
   2 & $N(\rm H_{2})  \; \rm cm^{-2}$    & $4.44 \times 10^{21}$   & $ 2.63 \times 10^{22}$  \\  
   3 & $A_{V}         \; \rm mag$	     & 2.0    		            & 12.0                    \\  
   4 & $N(\rm H_{2}O) \; \rm cm^{-2}$    & $ 2.0 \times 10^{17}$    & $ 1.0 \times 10^{18}$   \\ 
   5 & $N(\rm CO)     \; \rm cm^{-2}$    & -                        & $ 4.79 \times 10^{17}$  \\
\hline
\end{tabular}
\label{tab:table1_cond_core} \\
\end{table} 

\subsection{Mantle Compositions}
\label{sec:MC}
In a typical molecular cloud core condition, the surfaces of the bare grain substrates are expected to be covered by thick ice mantles, which may consist of up to a few hundred monolayers \citep{ormel2009dust,kalvans2015ice,chacon2019dust,caselli2022central}.

Kinetic chemistry models based on observational and experimental studies suggest that the structural characteristic of ice mantles are driven by the solid-phase formation/destruction efficiency of H$_2$O and the depletion/desorption level of CO on the grain surfaces \citep{brown1990chemical,watanabe2004hydrogenation, andersson2006molecular,garrod2008new,cuppen2017grain,iqbal2018statistical}.

According to the literature, H$_2$O- or CO-dominated mantles mainly originate from two main competitive ice formation processes on grains. The first is the effective H$_2$O mantles production via surface reactions beginning from the early chemical/dynamical state of interstellar medium (ISM) region known as the {\it early (or polar)} ice formation state, where the accretion of H- and O-rich atomic gas on grain surfaces is essential \citep{jones19843,boogert2011ice,oberg2011ices}. The second is the later accretion state of CO molecules over already formed water layers known as the CO freeze-out state that becomes {\it catastrophic (or apolar)} when an ISM region reaches a specific density ($n \sim$ a few $\mathrm{10}^{4}$  $\mathrm{cm}^{-3}$) and temperature ($T\simeq 10$ K) limits \citep{palumbo19932140,caselli1999co,jorgensen2005molecular,
pontoppidan2006spatial,pontoppidan2008c2d,cuppen2011co,qasim2018formation}. 

H$_2$O ice has a high sublimation point connected with its strong polar hydrogen bond networks, while CO ice has a lower surface binding energy due to the fact that CO mantle molecules bonded via weaker apolar van der Waals interactions \citep{fraser2004using}. To define H$_2$O- or CO-dominated mantles, we prefer the terms: polar and apolar as similarly used in some previous studies \citep[e.g.,][]{ehrenfreund1999laboratory,watanabe2004hydrogenation,cuppen2011co,gorai2020systematic}. 

For the reasons mentioned above, to simulate discrete mantle types that represent competitive and individual ice formation states within different conditions of the same molecular cloud core, we select two stratified pure ice mantles with layers consisting of either H$_2$O or CO.
\subsection{Icy Grain Model}
\label{sec:Icy Grain}
The basic properties of icy dust grains in our model are selected for the specific situations that correspond to the typical physical conditions of isolated molecular cloud core. To eliminate some computational complications upon the adopted dust grain model, we consider seven reasonable simplifying assumptions about the grain material and the size evolution, mainly based on the observational and the theoretical constraints. The results of the icy grain model are listed in Table \ref{tab:table2_MRN_grain_model}. We also give details of the icy grain model calculation in Appendix \ref{sec:appA}.
\begin{itemize}[left=0\parindent]
\item [1.] \textit{The total dust to gas mass ratio}: We assume the total dust to gas mass ratio ($R_{d}$) as 0.01, generally consistent with values derived from prominent dust models for diffuse ISM conditions\citep[e.g.,][]{li2001ultrasmall,li2001infrared, draine2007infrared}. The fiducial value of $R_{d}$ is still physically acceptable for a typical molecular cloud core because we only consider the formation of multiple ice layers with nearly a hundred-angstrom depths, which has quite limited effects on the grain
mass evolution \citep{ossenkopf1994dust,ormel2009dust,wada2009collisional,wettlaufer2010accretion,ormel2011dust,kohler2015dust,ysard2016mantle}. 
\item [2.] \textit{The chemical composition of bare grain}: Dust observations clearly show us that several sub-bare grain species with different chemical compositions exist. Mg/Fe rich silicates such as forsterite-type olivines are suitable candidates for the majority of astronomical grains \citep{sofia2001erratum,weingartner2001dust,compiegne2011global}. 
Therefore, we assume olivine as bare grain material, which has a bulk density of $\rho_{g}$ = 3.5 g/$\mathrm{cm}^3$ and typical chemical composition of MgFeSi$\mathrm{O}_4$  \citep{henning2010cosmic}.
\item [3.] \textit{The grain size distribution}: We take the standard MRN \citet{mathis1977size} grain size distribution function that follows a singular power-law with a -3.5 index.
The derivative form of this continuous distribution function can be simply defined as $\mathrm {dn_{dust} = C_{MRN} \; n_{H} \; a_{g}^{-3.5} \; da_{g}}$, where $n_{\rm dust}$ is the number density of dust grains with a specific size, ${\rm C_{\rm MRN}}$ is an integration constant, $n_{\rm H}$ is  the total gas hydrogen number density, and $a_{g}$ is the radius of fully spherical grain that has  a volume of $\rm V_{g} = \mathrm{4\pi a_{g}^{3}/3}$. 
According to the adopted grain size range between ${\rm a_{\rm gmin}}$ = 0.03 $\mu$m and ${\rm a_{\rm gmax}}$ = 0.3 $\mu$m, we find a value of ${\rm C_{\rm MRN}}$ = $ 3.579 \times 10^{-25} \rm cm^{2.5}$ for the silicate(olivine) grain.
\item [4.] \textit{The effective grain radius}: The MRN is not only a continuous grain size distribution but it also can be separated into size intervals \citep{pauly2016effects,sipila2020effect}. To evaluate the size-dependent ice formation limits in our dust grain model, we divide the  MRN distribution into ten size intervals. These intervals are equally and logarithmically dispersed throughout the full range of the grain cross-section [$\rm log \sigma_{gmin} -  \rm log\sigma_{gmax}$]. The calculated effective grain radii ($\rm a_{\rm gef}\;[\mu m]$) are given in Table \ref{tab:table2_MRN_grain_model}.
\begin{table}
\caption{The MRN grain size distribution results.}                 
\centering                      
\begin{tabular}{c c c c c c}      
\hline\hline            
\multicolumn{5}{c}{\textbf{EPS}}\\ 
\multicolumn{5}{c}{Ice mantle type$^{A1}$: thin H$_2$O}\\
\hline
$N_{interval}^{B}$ & $a_{gef}^{C}$ & ${\rm D_{\rm ice,max}}^{D}$ & ${\rm f_{\rm ice}}^{E}$ & ${\rm V_{\rm ratio}}^{F}$ & ${\rm T_{\rm surf}}^{G}$ \\   
\noalign{\smallskip}
 & $(\mu m)$ &  ($\AA$) & & &(K) \\
\hline
1  & 0.039  & 51 &  1.0  & 0.53 & 11.27 \\
2  & 0.047  & 45 &  1.0  & 0.36 & 10.91 \\
3  & 0.057  & 40 &  1.0  & 0.25 & 10.54 \\
4  & 0.070  & 36 &  1.0  & 0.17 & 10.18 \\
5  & 0.087  & 32 &  1.0  & 0.12 & 9.81  \\
6  & 0.109  & 28 &  1.0  & 0.08 & 9.45  \\
7  & 0.136  & 25 &  1.0  & 0.06 & 9.10  \\
8  & 0.170  & 22 &  1.0  & 0.04 & 8.77  \\
9  & 0.213  & 20 &  1.0  & 0.03 & 8.44  \\
10 & 0.267  & 18 &  1.0  & 0.02 & 8.12  \\
\hline\hline   
\multicolumn{5}{c}{\textbf{CPS}}\\ 
\multicolumn{5}{c}{Ice mantle type$^{A2}$: thick H$_2$O}\\ 
\hline  
$N_{interval}$ & $a_{gef}$ & ${\rm D_{\rm ice,max}}$ & ${\rm f_{\rm ice}}$ & ${\rm V_{\rm ratio}}$ & ${\rm T_{\rm surf}}$ \\    
 \hline
1  & 0.066  & 327  & 0.844  & 4.06 & 8.73  \\  
2  & 0.071  & 291  & 0.845  & 2.57 & 8.63  \\  
3  & 0.079  & 260  & 0.849  & 1.65 & 8.48  \\  
4  & 0.090  & 231  & 0.844  & 1.09 & 8.29  \\  
5  & 0.105  & 206  & 0.853  & 0.73 & 8.08  \\  
6  & 0.124  & 184  & 0.836  & 0.49 & 7.85  \\  
7  & 0.150  & 164  & 0.852  & 0.34 & 7.60  \\  
8  & 0.182  & 146  & 0.854  & 0.23 & 7.35  \\  
9  & 0.224  & 130  & 0.837  & 0.16 & 7.10  \\  
10 & 0.277  & 116  & 0.842  & 0.11 & 6.85  \\     
\hline\hline   
\multicolumn{5}{c}{\textbf{CAPS}}\\ 
\multicolumn{5}{c}{Ice mantle type$^{A3}$: thick H$_2$O +  CO}\\
\hline
$N_{interval}$ & $a_{gef}$ & ${\rm D_{\rm ice,max}}$ & ${\rm f_{\rm ice}}$ & ${\rm V_{\rm ratio}}$ & ${\rm T_{\rm surf}}$ \\   
\hline
1  & 0.078  & 450  & 0.267 & 0.66 & 8.49  \\ 
2  & 0.082  & 401  & 0.269 & 0.53 & 8.41  \\ 
3  & 0.089  & 357  & 0.269 & 0.42 & 8.31  \\ 
4  & 0.099  & 318  & 0.264 & 0.32 & 8.16  \\ 
5  & 0.112  & 283  & 0.264 & 0.24 & 7.98  \\ 
6  & 0.131  & 253  & 0.261 & 0.18 & 7.78  \\ 
7  & 0.156  & 225  & 0.266 & 0.13 & 7.55  \\ 
8  & 0.188  & 201  & 0.269 & 0.09 & 7.32  \\ 
9  & 0.229  & 179  & 0.268 & 0.07 & 7.07  \\ 
10 & 0.282  & 159  & 0.263 & 0.05 & 6.83  \\ 
\hline                       
\end{tabular}
\label{tab:table2_MRN_grain_model} \\
\begin{itemize}
\item[A:] A1 corresponds to pure H$_2$O ice mantle at the end of the edge polar state. A2  corresponds to pure H$_2$O ice mantle at the end of the center polar state that has a thicker mantle depth than A1. A3 corresponds to the ice mantle at the end of the center apolar state that consists of the outer CO  and the inner H$_2$O components.
\item[B :]The label of the grain size intervals.
\item[C :]The size-dependent effective grain radius at the end of the specific ice formation state.
\item[D :]The size-dependent ice mantle depth at the end of the specific ice formation state.
\item[E :]The mantle depth fraction of the recently accreted ice on the grain surface. $f_{ice}$ $\times$ ${\rm D_{\rm ice}}$ gives a mantle depth of newly formed ice at the end of the specific ice formation state.
\item[F :]The size-dependent volume increment factor at the end of the specific ice formation state.
\item[G :]The size-dependent grain surface temperature at the end of the specific ice formation state.
\end{itemize}
\end{table} 
\item [5.]\textit{The representative grain radius}: We define the total number of surface binding sites of the grain size distribution in the range $a_{\rm gmin} - a_{\rm gmax}$ as $\mathrm{\sum_{k = 1,10}[X_{d,k}\, 4\,\sigma_{gef,k}] \,l_{s}^{-2}}$.  $X_{\rm d,k}$ is the effective grain abundance in each size interval k (1 to 10).   $\sigma_{\rm gef,k}$ is the effective grain cross section in each size interval k (1 to 10).  In line with previous studies \citep [e.g.,] [] {herbst2005chemistry,cazaux2016dust,hocuk2016chemistry, pauly2016effects, zhao2018effect}, that suggest the binding sites are homogeneously distributed across the grain surface, we assume that $\rm l_{s} = 3 \, \AA$ is the typical distance between two binding sites on the grain (which is also the depth of one ice layer).
To derive the representative grain radius of the adopted distribution, we equalize the total number of surface binding sites of the distribution to a singular number of surface binding sites value of the grain with radius $\rm a_{\rm grep}$  and abundance  $X_{\rm d,rep}$.
According to this equalization, we find a value of $\rm a_{\rm grep}$ = 0.095 $\mu$m with the accompanied number of surface binding sites per $\rm cm^3$, which is equal to 3.95 $10^{-6}$ $n_{\rm H}$ $ \rm cm^{-3}$.
\item [6.] \textit{The ice mantle depth}: In evaluating the size-dependent ice mantle depths, we first use the approximation in \citet{kalvans2018temperature} that suggests an almost linear correlation between the mantle depth of a grain with a specific radius and the visual extinction of the medium. According to this approximation, we calculate the maximum mantle depth for the representative grain radius at the end of ice formation states as ${\mathrm{D_{\rm ice,max,a_{grep}}= d_{\rm ice,max} \times a_{\rm grep}}}$. $D_{\rm ice,max,a_{grep}}$ is the maximum mantle depth of the representative grain radius  at the end of three ice formation states whereas, ${\rm d_{\rm ice,max}}$ is a unitless constant.  To ensure physical consistency for the considered molecular cloud core condition, we choose a value of ${\rm d_{\rm ice,max}}$ = 0.305. This is proportional to the visual extinction obtained at the end of three ice formation states. Based on the adopted H$_2$O and CO ice column densities that correspond to the visual extinction values of $A_{\rm V,edge} = 2.0$ mag (for EPS), $A_{\rm V,middle} = 6.0$ mag (for CPS), and $A_{\rm V,center} = 12.0$ mag (for CAPS), we divide the ${\rm d_{\rm ice,max}}$ into three parts. These are ${\rm d_{\rm ice,a}}$ = 0.035, ${\rm d_{\rm ice,b}}$ = 0.187, and ${\rm d_{\rm ice,c}}$ = 0.083 for EPS, CPS, and CAPS, respectively. Considering this, we derive the mantle formation dependent volume increment factor, ${\rm V_{\rm ratio}}$ (the ratio of the mantle volume to the grain core volume), for each grain size interval at EPS, CPS, and CAPS (see Table \ref{tab:table2_MRN_grain_model}).
\item [7.] \textit{The grain surface temperature}: To calculate the effective surface temperatures of grains in EPS, CPS, and CAPS with respect to the adopted size distribution, we use $A_{V}$-dependent dust temperature expression in \citet{hocuk2017parameterizing}. Since this analytical expression gives the thermal equilibrium temperature of grain with the canonical radius of 0.1 $\mu$m, we apply a scaling procedure to derive the effective grain temperatures that correspond to the effective final gain radii at the end of the specific ice formation state. The adopted temperature scale for different grain sizes varies as $\rm a_{gef}^{-1/5.9}$. Using interstellar radiation field strength, $\rm G_{0}$ =1.31 \citep{marsh2014properties} in units of the Habing field \citep{habing1968interstellar}, we calculate the effective grain surface temperatures, ${\rm T_{\rm surf}}$ in unit of Kelvin (see Table \ref{tab:table2_MRN_grain_model}).
\end{itemize}
\subsection{The Stopping Power Data}
\label{sec:SPD}
The stopping power has two constituents that are the {\it nuclear} (or the knock-on) stopping power 
(hereafter ${\rm S_{\rm n}}$) and the {\it electronic} stopping power (hereafter ${\rm S_{\rm e}}$). ${\rm S_{\rm n}}$ is related to the atomic displacement cascades induced by the elastic collisions \citep{tolstikhina2018basic}. 
However, ${\rm S_{\rm e}}$ is driven by successive inelastic interactions such as ionization, excitation, and electron-phonon coupling \citep{agullo2005lattice}.
\begin{figure*} 
\resizebox{0.94\textwidth}{!}{
\centering
\includegraphics[scale=0.48]{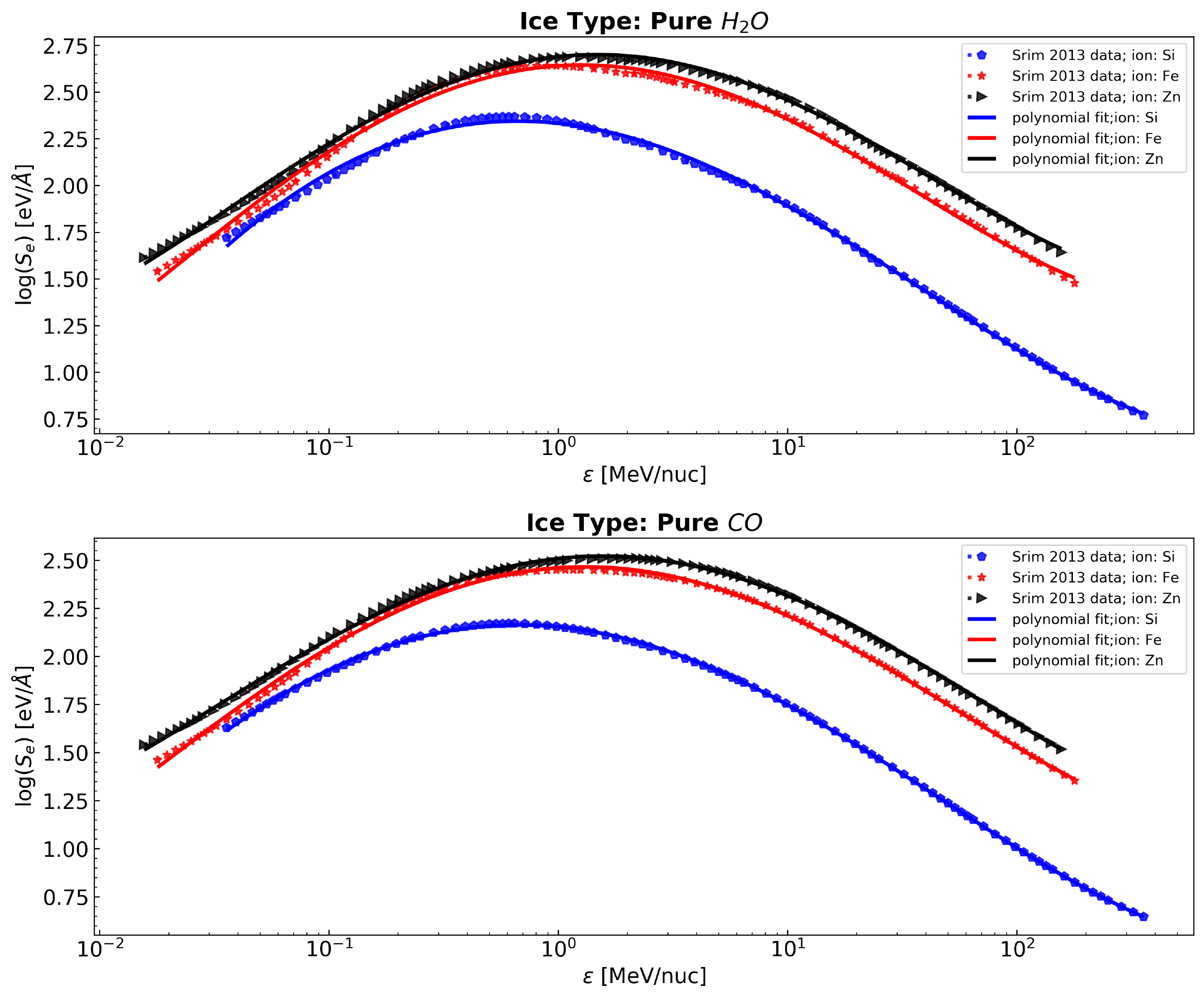}}
\caption{The electronic stopping power, ${\rm S_{\rm e}}$ (in units of eV/$\AA$) data  (loosely dotted lines) are derived from the SRIM 2013 package. The calculated fourth-degree polynomial ${\rm S_{\rm e}}$ fits (solid lines) as a function of the incident CR ion kinetic energy, $\varepsilon$ (in units of Mev/nuc). The SRIM 2013 data and the fitted polynomials are derived for pure H$_2$O and pure CO ice mantle compositions and three CR ion types (Si, Fe, Zn). }
\label{fig:fig2_electronic_stp}
\end{figure*}
In obtaining the stopping power data of both H$_2$O and CO ice in the logarithmic initial kinetic energy range 1 MeV – 10 GeV that consists of 105 data points, we use the SRIM 2013 package \citep{ziegler2010srim} for 30 CR ion types, including both light (Z = 1, 2) and heavy (Z $\geq$ 3) atomic numbers from proton to Zn. It is worth noting that, during the stopping power calculations, we assume the Bragg correction coefficient \citep{powers1980influence,ziegler2004srim} is equal to 1 for two ice components.

Since the electronic stopping regime ultimately governs the CR-induced hot spot heating process \citep{johnson2013sputtering}, we only consider ${\rm S_{\rm e}}$ data values in this study. 

In Figure \ref{fig:fig2_electronic_stp}, the dotted lines show the derived ${\rm S_{\rm e}}$ data as a function of CR ion kinetic energy per nucleon. The solid lines correspond to the fourth-degree polynomial fits to each specific ${\rm S_{\rm e}}$ data set. 
\subsection{Elemental Composition and Energy Spectrum  of  CR Ions}
\label{sec:CRSPEC}
We adopt relative elemental compositions (or abundances) of each 30 different CR ion species in table 2 of \citet{kalvans2018temperature}. The adopted fractional CR abundances correspond to local Galactic CR abundances, which are based on mostly the data from Voyager I \citep{webber1983measurement} and the other measurements of many space-borne experiments, e.g., CRIS \citep{stone1998cosmic}, PAMELA\citep{orsi2007pamela}, INTEGRAL \citep{tatischeff2012nonthermal}, and SUPERTIGER \citep{binns2014supertiger} as mentioned in previous studies \citep{shen2004cosmic,chabot2016cosmic,kalvans2018temperature}.

Based on the leaky box model \citep{ip1985estimates}, the isotropic local energy spectrum (or flux) of proton CRs is described as the number of protium particles within a definite energy range per unit area, per solid angle, per unit time, and per atomic mass unit. This energy spectrum based on space-probe measurements is  represented by power-law distribution functions for high energy regime (1 GeV/nuc) \citep{gabici2019origin}. However, a reliable analytical expression of the low energy part of the spectrum might be difficult to achieve because of the strong solar modulation effects on the measurements induced by attenuating interactions between the material in the solar wind and proton CRs with relatively low energies \citep{takayanagi1973molecule,padovani2009cosmic}.

\begin{figure*}
\resizebox{0.95\textwidth}{!}{
\centering
\includegraphics[scale=0.3]{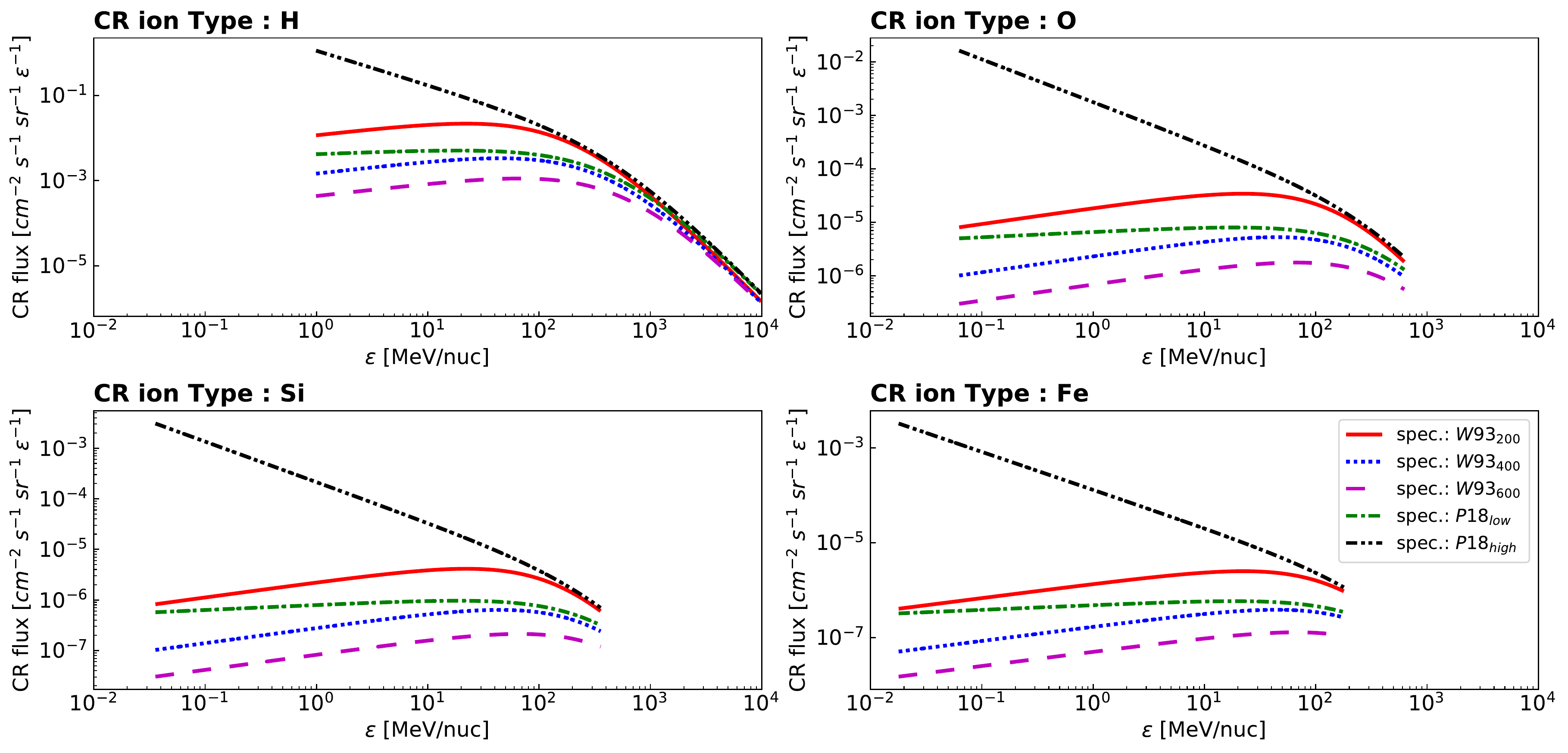}}
\caption{The abundance-dependent differential energy distribution of four different CR ions (H, O ,Si and Fe) as a function of CR kinetic energy per nucleon in the range $10^{-2} - 10^{4}$ MeV/nuc. To consider the effects of low-energy CR ions in the spectrum, we use five CR ion fluxes that consist of three W93 \citep{webber1983measurement} and two P18 \citep{padovani2018cosmic} energy distribution functions with different low-energy contents. W93 spectra with different form parameters ($\rm E_{\rm C}$), which specify the number of low-energy CRs, are shown as: red solid line for $\rm E_{\rm C}$ = 200 MeV, blue dotted line for $\rm E_{\rm C}$ = 400 MeV, magenta dashed line for $\rm E_{\rm C}$ = 600 MeV. Whereas P18 spectra are shown as: green dash-dotted line for $\rm P18_{\rm low}$ and lack double dash-dotted line for $\rm P18_{\rm high}$.}
\label{fig:fig3_CR_spectra}
\end{figure*}

The characteristic form of proton CR spectrum can be dramatically influenced by the number of low-energy CRs, which may exhibits distinct variations depending on local conditions             \citep{padovani2018cosmic,silsbee2021ice}. These variations are very critical for the alteration of CR ion energy deposition efficiency into grain surfaces \citep{dartois2013swift,dartois2021cosmic}. 
To consider the influence of the low-energy part of the proton CR flux, we use two functional CR energy distribution approximations and five different power-law functions with the same characteristic functional complexions. Figure \ref{fig:fig3_CR_spectra} shows the adopted five functions with specific coefficients, which constrain the effect of low energy CR contents on the spectrum. Two of the adopted CR energy spectrum functions are obtained from \textit{the high} and \textit{the low} energy spectra (hereafter $\rm P18_{\rm low}$ and $\rm P18_{\rm high}$), developed by \citet{padovani2018cosmic}, whereas the other three functions are derived from the spectrum (hereafter $W93$) of \citet{webber1983measurement} that has been used before in many studies  \citep[e.g.,][] {shen2004cosmic,dartois2013swift}.
Assuming that the local energy distributions of both light and heavy CR ions are almost the same as proton CRs for the adopted cloud core conditions \citep{chabot2016cosmic}, we individually obtained fluxes of other 29 CR ion components in our dataset by multiplying the proton CR energy spectrum with the relative elemental abundance of relevant CR ions. 
The flux for a specific CR ion type as a function of  energy per nucleon (or per atomic mass unit) in our model is 
\begin{equation}
\label{eq:Equation 1}
j_{A_i\;>1}(\varepsilon_i) = \rm f_{Z_{CRi} \;> 1} \times\; j_{p}(\varepsilon_i), \;\;cm^{-2} s^{-1} sr^{-1} \varepsilon^{-1},
\, 
\end{equation}   
where $\varepsilon_i = E_i/A_i$ is the kinetic energy per nucleon of CR ion i, $E_i$ is the kinetic energy, $A_{i}$ the atomic mass number, $\rm f_{Z_{CRi} \;> 1}$ is the fractional elemental abundance of the relevant CR ion that has  $Z_{i}$ proton number. $j_{p}$ is the energy spectrum of CR proton.
\subsection{The Analytical Thermal Spike Model}
\label{sec:ATS}
To determine the thermal spike radius (latent track radius) induced by the incident CR ions with a broad energy range within ice mantles on grain surfaces, we choose the analytical thermal spike model proposed by  \citet{szenes1997amorphous} because of three reasons. First, the analytical thermal spike model (hereafter ATS) does not consider the actual time evolution of the spot heating process. The ATS model assumes that the CR-induced temperature increase on a target material can be approximated by the Gaussian distribution function, which depends on the volumetric heat capacity and the ${\rm S_{\rm e}}$  \citep{wesch2016ion}. Second, the maximum width of temperature distribution within the latent region and the characteristic cylindrical shape of this region around the CR ion path can be insensitively defined to the material properties such as heat conduction, bandgap energy, and chemical composition, and degree of crystallization \citep{szenes1996formation}. Third, the validity of the ATS model is clearly confirmed by many empirical studies of quite different materials, e.g., semiconductors, magnetic insulators, $\alpha$-quartz, and mica \citep{szenes2002tracks,szenes2011comparison}.

According to the ATS model, to evaluate the maximum (or initial) radii of CR-induced cylindrical latent regions on icy grain surfaces as a function of ${\rm S_{\rm e}}$, we use five assumptions explained as follows:
\begin{itemize}[left=0\parindent]
\item [1.] The maximum (or the initial) width of the Gaussian local temperature distribution at  $t = 0$ s  within the latent track, a(0) is fixed and equals 45 {\AA} and it also is independent of ice mantle composition.  Therefore, we use the same a(0) value for both H$_2$O and CO ice.
\item [2.]A definite threshold electronic stopping power value (hereafter ${\rm S_{\rm e,thres}}$) exists to form the latent track with a continuous and cylindrical shape. This threshold is proportional to four parameters:  a(0),  the specific heat of ice, ${\rm c_{\rm ice}}$,  the bulk density of ice material, ${\rm \rho_{\rm ice}}$, and the transient CR induced temperature increment within the latent track, ${\rm \Delta}\rm T_{\rm 0}(\rm R_{\rm max,track}, \ t= 0 \ s)$, which is calculated by the difference between the melting point temperature of ice, ${\rm T_{\rm melting}}$  and the initial effective surface temperature in each size interval before CR irradiation, ${\rm T_{\rm initial}}$ (which is equal to the effective grain surface temperature, ${\rm T_{\rm surf}}$).  
\item [3.] ${\rm S_{\rm e}}$ induced energy deposition fraction on the mantle is mainly driven by two parameters: swift secondary electrons losses from the mantle, 1 - $\alpha$ \citep{leger1985desorption} and the efficiency of electron-phonon coupling within the latent track, $\gamma$ \citep{wesch2016ion}. The maximum latent track radius (${\rm R_{\rm max,track}}$) and ${\rm S_{\rm e,thres}}$ values are very sensitive to the fraction of locally deposited energy on icy grain surfaces \citep{baragiola2003sputtering} due to the fact that the energy deposition fraction on the mantle is reduced by $\alpha \times \gamma$.
\item [4.] To calculate melting point temperatures of pure H$_2$O and CO ices, we use a simplified linear approximation derived from \citet{bringa2004new} as ${\rm T_{\rm melting}} \approx (0.1 \times {\rm E_{\rm bind,ice}})$. Where  ${\rm E_{\rm bind,ice}}$ is the ice surface binding energy. In line with table A.1 of \citet{hocuk2016chemistry}, we adopt ${\rm E_{\rm bind,ice}}$ = 5700 K for H$_2$O and ${\rm E_{\rm bind,ice}}$ = 1300 K for CO .
\item [5.] The volumetric heat capacities, ${\rm C_{\rm V,ice}}$ of H$_2$O and CO ices can be evaluated by multiplying the bulk density and the specific heat as ${\rm \rho_{\rm ice}}\times {\rm c_{\rm ice}}$. To consider a more realistic grain mantle structure with an amorphous shape, which probably consists of many micro-pores, we adopt bulk densities used by \citet{dartois2015heavy,dartois2021cosmic}  for pure H$_2$O and CO ices: as $\rho_{\rm H_2O}$ = 0.93 $g/cm^{3}$ and $\rho_{CO}$ = 0.8 $g/cm^{3}$.
Since the surface temperature sensitivity of the ice specific heat is almost negligible at high-temperature regimes like the melting point (see \citet{schmalzl2014earliest} for details), we choose Neumann Kopp’s rule for the ice-specific heat calculation method rather than using a complex temperature-dependent approach.  

Our specific heat approximation is   
\begin{equation}\label{eq:Equation 2}
\mathrm{\rm c_{ice}} =  {\rm \frac {3  N_{A} k_{B}}  {M_{ice} N_{atom}} \; J g^{-1} K^{-1}.
}\,
\end{equation} 
where ${\rm N_{\rm A}}$ is the Avogadro number, ${\rm k_{\rm B}}$ is the Boltzmann constant, ${\rm M_{\rm ice}}$ is the molar mass of the ice (18.015 for H$_2$O and 28.01 for CO), and ${\rm N_{\rm atom}}$ is the number of atomic components within an ice molecule (${\rm N_{\rm atom}}$ = 3 for H$_2$O   and  ${\rm N_{\rm atom}}$ = 2 for CO).
We calculate ice-specific heats from Equation (\ref{eq:Equation 2}) as 4.154 $\rm {J \, g^{-1}\, K^{-1}}$ and 1.781 $\rm {J \, g^{-1}\, K^{-1}}$ for H$_2$O and CO ices, respectively.
\end{itemize} 
Under these assumptions, we evaluate ${\rm R_{\rm max,track}}$ considering two conditional functions that depend on ${\rm S_{\rm e,thres}}$ and ${\rm S_{\rm e}}$
\begin{equation}\label{eq:Equation 3}
      \mathrm{\rm R_{max, track}} =
         {\rm \sqrt{[a^{2}(0) \times ln(S_{e, ratio})]}\; ; \; S_{e,thres} \leq S_e \leq 2.7\,S_{e,thres},
}\,
   \end{equation}   
\begin{equation}\label{eq:Equation 4}
      \mathrm{\rm R_{max,  track}} =
         {\rm \sqrt{[(\frac{a^{2}(0)}{2.7})  \times (S_{e,ratio})]} \; ;\;  (S_e > 2.7\,S_{e,thres}),
}\,
   \end{equation} 
\begin{equation}\label{eq:Equation 5}
      \mathrm{\rm S_{e,  ratio}} =
         {\rm \frac{S_e}{S_{e,thres}}, 
}\,
   \end{equation}   
\begin{equation}\label{eq:Equation 6}
      \mathrm{\rm S_{e, thres}} =
         {\rm \frac{C_{V \,ice} \times \sigma_0 \times \Delta_{T_0} }{\alpha \times \gamma } \;\; \frac{eV}{\AA},
}\,
   \end{equation}     
\begin{equation}\label{eq:Equation 7}
      \mathrm{\rm \sigma_{0}} ={\rm \pi \times a^{2}(0) \;\; \AA^2.
}\,
   \end{equation} 
\subsection{Hot spot Induced Sputtering}
\label{sec:HSIS}
In the sputtering yield calculation, we consider two sputtering sub-regimes,which correspond to the linear and the quadratic sputtering as functions of ${\rm S_{\rm e}}$.  

In describing the transitional yield evolution from the linear to the quadratic regime, we use the latent track formation thresholds that are governed by the effective electronic stopping power (hearafter ${\rm S_{\rm e,eff}}$). According to our assumption, the sputtering yields vary quadratically when the latent track formation is allowed (${\rm S_{\rm e,eff}}  > {\rm S_{\rm e,thres}}$), whereas yields show a linear dependence for the ${\rm S_{\rm e,eff}}$ values below the latent track formation limits (${\rm S_{\rm e,eff}}  < {\rm S_{\rm e,thres}}$).
To calculate yields for two sputtering sub-regimes more precisely, we make five corrections, which  have critically affect results. These corrections are given as follows:
\begin{itemize}[left=0\parindent]
\item [1.] \textit{The effective stopping power}: After a CR ion-icy grain collision, the incident CR ion energy is mainly deposited on the electronic subsystem of the radiated mantle material depending on the ${\rm S_{\rm e,eff}}$. Since the number of charge carriers and the ability of electron-phonon coupling are crucial in specifying how efficient the partly conversation of electronic energy into thermal energy \citep{toulemonde2004track,szenes2011comparison}, ${\rm S_{\rm e}}$ values should be corrected by multiplying with $\alpha$ and $\gamma$ factors to calculate ${\rm S_{\rm e,eff}}$ values. Here the $\alpha$ is a reduction factor for secondary ($\delta$) electrons-induced energy lost from the grain mantle, and the $\gamma$ is the electron-phonon coupling efficiency constant, which gives the deposited thermal energy fraction within the atomic sub-system of the mantle.
\item [2.] \textit{The impact angle}: In calculating the sputtering yield increment factor, $\zeta$ as a function of  $\rm \theta$ and the grain size, we assume three cases according to the studies of  \citeauthor{leger1985desorption, bringa2001angular}\,(\citeyear{leger1985desorption, bringa2001angular}; hereafter L85 and B01, respectively). Where $\rm \theta$ is the incident angle of CR ion.
In the first case, $\zeta$ shows relatively smooth variations that scale with $[{cos \theta}]^{-1}$ whereas, in the second and third cases, $\zeta$ rises more steeply depending on [$\mathrm{cos \theta}]^{-1.6}$ and on [$\mathrm{cos \theta}]^{-1.7}$, respectively. In the quadratic sputtering regime, we take the first two cases for $\zeta$ factor, while in the linear sputtering regime, we only consider the third case for the variation of $\zeta$ factor (see Appendix  \ref{sec:appB} for details).
\item [3.] \textit{The intersect points}: In our calculation parameter space, the stopping ranges of CR ions are much bigger than the total grain diameters for all the size bins. Hence, in taking into account the sputtering from the CR ion-grain intersect points at both sides of the ice mantle, we correct the sputtering yields by multiplying by a factor of two. 
\item [4.] \textit{The exponential decay}: \citeauthor{dartois2018cosmic,dartois2020electronic, dartois2021cosmic}\,(\citeyear{dartois2018cosmic,dartois2020electronic, dartois2021cosmic}; hereafter D18, D20 and D21, respectively) suggest that an explicit dependency between the sputtering yield and the ice mantle depth exists for different ice materials. According to this suggestion, sputtering yields can be noticeably decreased when the characteristic probe depth of the ice mantle (${\rm D_{\rm p,ice}}$)  exceeds the maximum mantle depth (${\rm D_{\rm ice,max}}$). ${\rm D_{\rm p,ice}}$ varies depending on ${\rm S_{\rm e,eff}}$ values. This parameter corresponds to a specific maximum depth where the deposited energy within the mantle mostly contributes to the sputtering. Therefore, to consider the reduction of yields for thin ice mantle situations ($D_{p,ice} \geq {\rm D_{\rm ice,max}}$), yields should be corrected by an exponential decay factor that is defined as 
$\mathrm{\rm n(\chi) = [1 - e^{-\chi}]}$.  Here $\rm \chi$ is the ratio of ${\rm D_{\rm ice,max}}$ to ${\rm D_{\rm p,ice}}$. In calculating ${\rm D_{\rm p,ice}}$ values for H$_2$O and CO ices, we use two equations that are 
\begin{equation}\label{eq:Equation 8}
      \mathrm{N_{p,ice}} =  10^{r} \times (\rm S_{e, eff} \times \rm \beta)^{s} \; \rm cm^{-2},
         {
}\,
   \end{equation}
\begin{equation}\label{eq:Equation 9}
      \mathrm{D_{p,ice} = \Bigg[\frac{N_{p,ice} \times M_{ice}}{\rho_{ice} \times N_{A}}\Bigg]} \times 10^{8} \; \AA,
         {
}\,
   \end{equation}   
where $N_{p,ice}$ is the ice column density (in units of $cm^{-2}$) at the probe depth, $\beta$ is the constant for the unit conversion (\rm $eV/\AA$ to $eV/10^{15}$ molecules/$cm^{2}$) of ${\rm S_{\rm e,eff}}$ values. The power-law indexes (r and s) in Equation (\ref{eq:Equation 8}) are derived from D18 and D21 for H$_2$O and CO ices, respectively. The adopted values of power-law indexes are $\rm r_{H_2O}$ = 14.011, $\rm r_{CO}$ = 14.25 and $\rm s_{H_2O}$ = 1, $\rm s_{CO}$ = 0.95.
\item [5.] \textit{The moving direction of sputtering}: In considering the sputtered ice molecules that are forwardly directed toward the mantle surface, we use a correction factor defined as $\Upsilon$.  Based on the molecular dynamics calculations of \citet{bringa2004new,johnson2013sputtering}, we adopt $\Upsilon$ = 0.1 for H$_2$O and CO ices in both sputtering regimes. 
\end{itemize}
\subsubsection{Sputtering Yields}
\label{sec:SY}
In the quadratic sputtering regime, where the latent track formation threshold is surpassed, we assume two energy distributions for thermalized ice molecules within the cylindrical track region: the Maxwellian distribution and the non-Maxwellian distribution (i.e., a $\delta$ distribution). \citeauthor{johnson1991linear}\,(\citeyear{johnson1991linear}; hereafter J91) suggest that the energy distribution characteristic is controlled by a conditional function defined as g(1/$\xi$). 

For H$_2$O and CO ice mantles, $\xi$ can be taken as the ratio of the electronic excitation density inside the latent track to the surface binding energy density of the ice mantle. 
We take $\rm \xi$ = [$S_{\rm e, eff}$ / $\pi$  $R_{\rm max, track}^{2}$] / [$\rm l_{s}^{-3} \times \rm E_{\rm bind, ice}$].

According to J91, the energy distribution becomes the Maxwellian for 1/$\xi$ $\gg$ 1 case, whereas the $\delta$ distribution is only valid for 1/$\xi$ $\ll$ 1 case. 
To evaluate g(1/$\xi$) function for the Maxwellian distribution and the non-Maxwellian cases, we take the analytical  solutions that are given in appendix A.2 of \citet{bringa1999molecular}.  

Considering these assumptions, we rearrange the quadratic yield equation for H$_2$O and CO  ices  as follows
\begin{equation}\label{eq:Equation 10}
      \mathrm{\rm Y_{hs}^{qu}} = 2 \times \overline{\zeta} \times \Gamma \;\Bigg[\frac{\rm S_{e,eff} \times l_{s}\times \Upsilon }{\rm E_{bind, ice}}\Bigg]^{2} \times g(1/\xi) \times n(\chi),
         {
}\,
   \end{equation}
where $\Gamma$ is a proportionality constant that is related to the energy distribution type. We adopt 
$\Gamma$ values of J91 that equals 1 and 0.4 for the Maxwellian and the $\delta$ distributions, respectively.
To ensure the continuity of results during the transition between two extreme cases (1/$\xi$ $\gg$ 1 and 1/$\xi$ $\ll$ 1), we apply a smoothening approximation to the quadratic yield expression: 
\begin{equation}\label{eq:Equation 11}
      \mathrm{f(1/\xi) = \frac{1 + tanh[(10^{p} \times 1/\xi) - 1]}{2}}, 
         {
}\,
   \end{equation}  
\begin{equation}\label{eq:Equation 12}
\mathrm{\rm Y_{hs}^{qu}} = \rm Y_{hs,\delta}^{qu} + \Big[\rm f(1/\xi) \times \big(\rm Y_{hs,Maxwellian}^{qu} - \rm Y_{hs,\delta}^{qu}\big) \Big].
\end{equation} 
We choose 0.18 and 0.398 as the smoothening degree values of the power-law index p in Equation (\ref{eq:Equation 11}) for H$_2$O and CO ices.
In the linear sputtering regime where the cylindrical latent track formation is not allowed, we use a linear approximation for the sputtering yield: 
\begin{equation}\label{eq:Equation 13}
      \mathrm{\rm Y_{hs}^{li}} = 2 \times \overline{\zeta} \times 0.14\;\Bigg[\frac{\rm S_{e,eff} \times l_{s} \times \Upsilon
      }{\rm E_{bind, ice}}\Bigg] \times \rm n(\chi).
         {
}\,
   \end{equation}
In evaluating the sputtering fluxes (in units of molecules$/\rm cm^{2}/ \rm s$) for H$_2$O and CO ices according to 30 CR ion types with different abundances, we first multiply $\rm \varepsilon$-dependent sputtering yields in two sputtering regimes with abundance-dependent differential CR ion flux. We then separately integrate these products over the $\rm \varepsilon_{\rm i}$ ranges for each CR ion. The cumulative sputtering flux, $\rm F_{\rm ice}$ equals the summing of 30 integral results for two sputtering regimes:
\begin{equation}
\label{eq:Equation 14}
\mathrm{\rm F_{ice}^{qu,li}} = 4\pi \times \rm {\sum}_{Z_i}\Bigg[\int_{\rm \varepsilon_{i,min}}^{\rm \varepsilon_{i,max}} {\rm Y_{hs}^{qu,\; li}(\rm \varepsilon_i) \times j_{\rm \varepsilon_i} \; \rm d\varepsilon \Bigg]\;    
}\, \;\frac{\rm molecules}{\rm cm^{2} \times \rm s}.
   \end{equation} 
Multiplying $\rm F_{\rm ice}$ values by a constant, $\rm q_{MRN}$, we obtain sputtering rate coefficients  of H$_2$O and CO ice molecules for ten grain size bins at three separate ice formations states. The $\rm q_{MRN}$ is the ratio of the effective grain area $(\rm \sigma_{\rm gef,k} \times X_{\rm d,k})$  to the number of surface binding sites ($\rm N_{\rm site,gef,k}$) for each grain size interval k (1 to 10).  We find the $\rm q_{MRN}$ constant as $\mathrm{2.5 \times 10^{16} \; cm^{2}}$. Using this constant, we eventually derive the analytical form of the sputtering rate coefficient, ${\rm k_{\rm hs}}$, (in unit of molecules per second).  ${\rm k_{\rm hs}}$ can be written as 
\begin{equation}\label{eq:Equation 15}
      \mathrm{\rm k_{hs}^{qu,li}} = \rm q_{MRN} \times \rm F_{ice}^{qu,li} \; \;\frac{\rm molecules}{s}.
         {
}\,
   \end{equation}
\section{RESULTS and DISCUSSIONS}
\label{sec:DISCUSS}
\begin{figure*}
\resizebox{0.96\textwidth}{!}{
\centering
\includegraphics[scale=0.48]{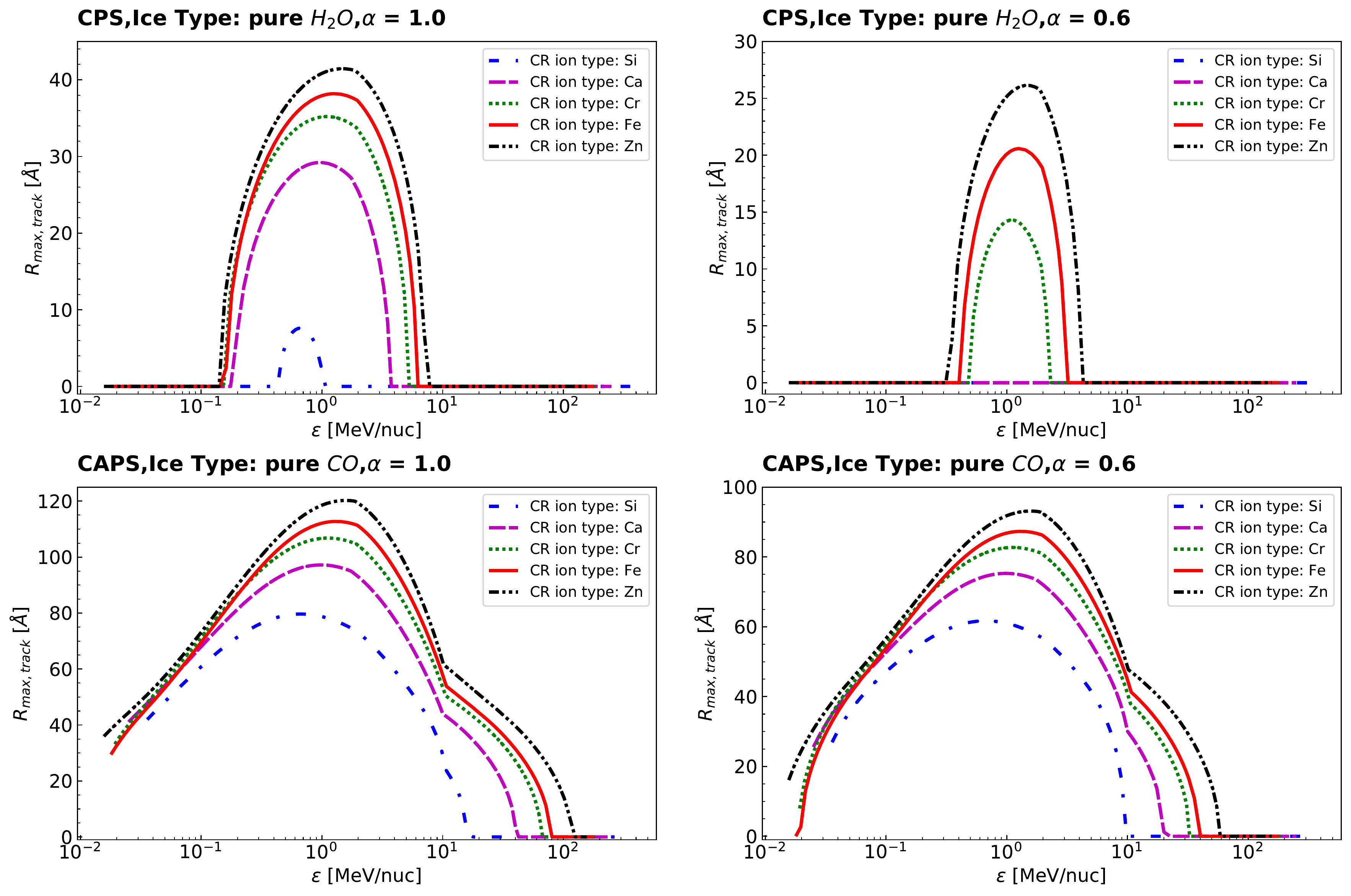}}
\caption{The maximum radius of the cylindrical latent radius, ${\rm R_{\rm max,track}}$ (in units of $\AA$) versus the incident CR ion kinetic energy (in units of MeV/nuc) for pure H$_2$O and CO ice mantles at two ice formation states (CPS and CAPS). 
The evolution of ${\rm R_{\rm max,track}}$ is controlled by the mantle composition, $\alpha$ reduction parameter, and the CR ion type. To represent the evolution characteristic, we show four situations that contain two $ \rm \alpha$ parameters ($\rm \alpha_{\rm max}$ = 1.0 and $\rm \alpha_{\rm min}$= 0.6) and five CR ion types (Si, Ca, Cr, Fe, and Zn) with the atomic numbers 14, 40, 52, 56 and 63, respectively. }
\label{fig:fig4_Rmax_variation}
\end{figure*} 
In this section, we discuss the  results obtained following  the calculation routines in Section ~\ref{sec:MM}.
\subsection{The Latent Track Formation}
\label{sec:LTF}
We find that the latent track formation process is directly linked to ${\rm S_{\rm e,eff}}$, which scales with  
$\approx \rm Z_{i}^{2}$. ${\rm R_{\rm max,track}}$ increases with the atomic number of the incident CR ion (see Figure \ref{fig:fig4_Rmax_variation}).  Therefore, the heavy CR ions are more suitable for the effective track formation on mantles. As seen in Figure \ref{fig:fig4_Rmax_variation}, the calculated ${\rm R_{\rm max,track}}$ radii for CO ice are much larger than the values of H$_2$O ice, and the relatively light ions may lead to the track formation on the CO ice. Contrary to the narrower profiles of H$_2$O tracks, CO track formations are allowed within the broader $\rm \varepsilon$ ranges.

We argue that the maximum radius of the cylindrical latent track radius, ${\rm R_{\rm max,track}}$ is quite sensitive to the ice mantle binding energy as well as the CR ion kinetic- energy-dependent effective electronic stopping power, $\rm S_{e,eff}(\rm \varepsilon)$. For this reason, the ice mantle composition and CR ion type play critical roles for ${\rm R_{\rm max,track}}$ evolution. We also find that ${\rm R_{\rm max,track}}$ shows variations according to the size-dependent initial grain surface temperature. However, the obtained correlation between ${\rm R_{\rm max,track}}$ and the grain size is relatively weak because  the effect of a high-temperature increment within the latent track at the ice melting point overcomes the size-dependent variation on the grain's initial surface temperature.

We find that ${\rm R_{\rm max,track}}$, scaled with $\rm S_{\rm e,eff}(\rm \varepsilon)$, tends to rise with the increase of the CR ion atomic number for both H$_2$O and CO ices. Therefore, heavy CR ions are more suitable inducers for latent track formation according to the ATS model. However, CO ice has larger ${\rm R_{\rm max,track}}$ values with respect to H$_2$O ice even for the same CR ion type and CR kinetic energy range, as expected. 

When ${\rm R_{\rm max,track}}$ values of the two mantle types are compared, it is clearly seen that the shapes of ${\rm R_{\rm max,track}}$ profiles are quite different. The characteristic ${\rm R_{\rm max,track}}$ profile of H$_2$O ice is narrower than the CO ice profile, which means that the latent track formation conditions are more restricted for H$_2$O ice with respect to the CO ice. 

In the case of $\alpha$ = 1, even light CR ions with $Z_{min} \geq 3$ can produce a cylindrical latent track within CO ice, whereas the cylindrical track formation within H$_2$O ice is only possible for heavier CR ion types with $Z_{min} \geq 14$. When the $\alpha$ is reduced to 0.6, the minimum CR ion atomic number thresholds that are needed for the cylindrical latent track formation increase depending on ice mantle chemical composition. In the case of $\alpha$ = 0.6, we find that the increased formation thresholds are  $Z_{min} \geq 22$ and $Z_{min} \geq 4$ for H$_2$O and CO ices, respectively. Therefore, we suggest that the effect of $\alpha$ reduction is more significant for H$_2$O ice. 
\begin{figure*}
\resizebox{0.96\textwidth}{!}{
\centering
\includegraphics[scale=0.48]{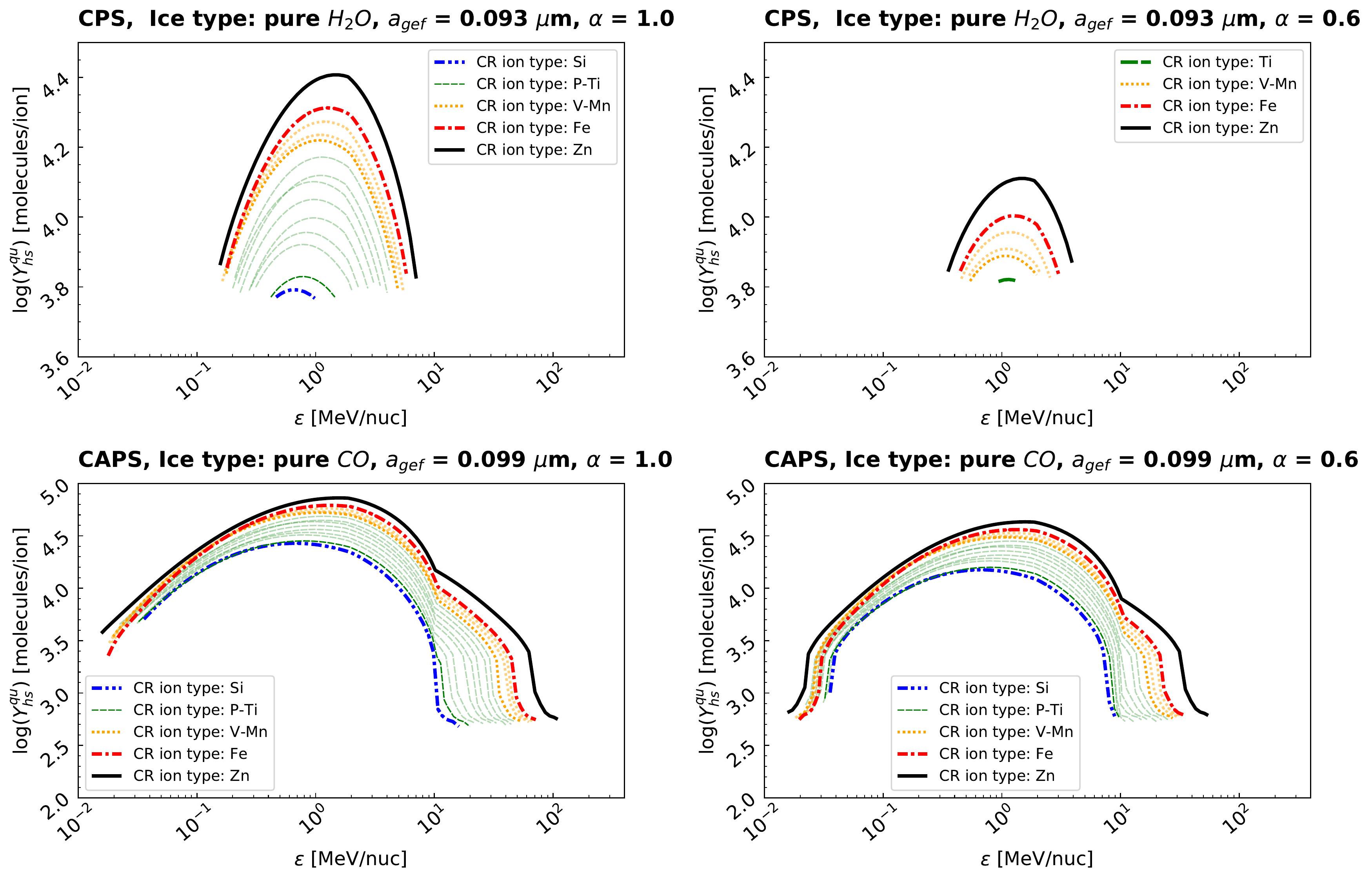}}
\caption{The quadratic sputtering yield, $\rm Y_{\rm hs}^{\rm qu}$ derived from Equation (\ref{eq:Equation 12}) as a function of the incident CR ion kinetic energy (in units of $\rm \varepsilon$) for pure H$_2$O and CO mantles at two ice formation states (CPS and CAPS).  $\alpha$ = 1 and $\alpha$ = 0.6 are reduction parameters on the ${\rm S_{\rm e}}$ values for $\delta$ electrons energy loss process.  $\rm a_{\rm gef}$ = 0.093 $\mu$m  and $\rm a_{\rm gef}$ = 0.099 $\mu$m are effective grain radii for the fourth grain size bin of the adopted MRN size distribution at the end of CPS and CAPS.
$\rm Y_{\rm hs}^{\rm qu}$  and $\rm \varepsilon$ axes are in logarithmic scales for the four panels. The legends show the CR ion types scaled with the proton number (blue dash-dotted line: Si, green dashed lines: P; S; Cl; Ar; K; Ca; Sc; Ti, orange dotted lines: V; Cr; Mn, red dash-dotted line: Fe and black solid line: Zn).}
\label{fig:fig5_quad_yiled_variation}
\end{figure*} 
\begin{figure*}
\resizebox{0.96\textwidth}{!}{
\centering
\includegraphics[scale=0.48]{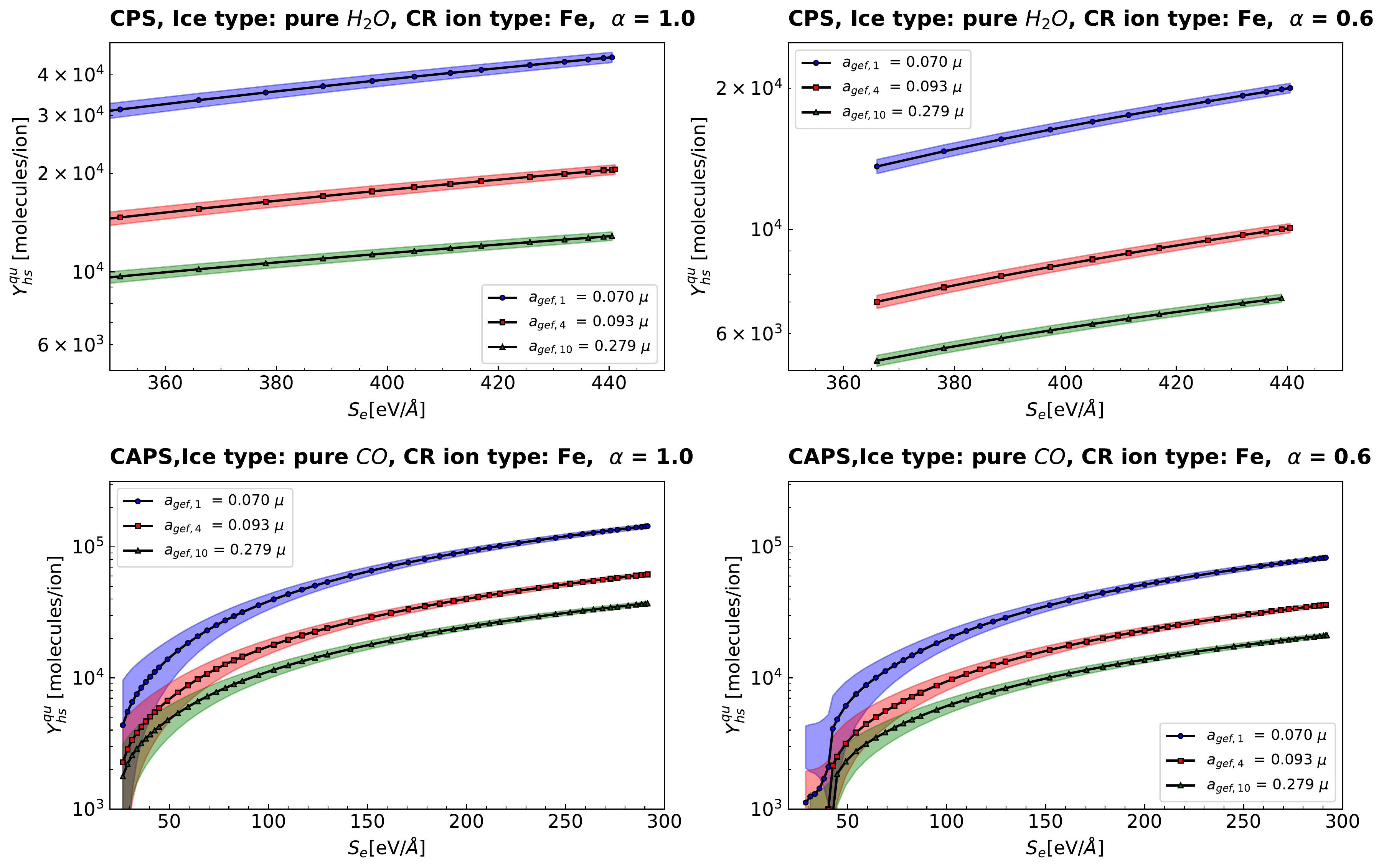}}
\caption{The grain size-dependent quadratic sputtering yields, $\rm Y_{\rm hs}^{\rm qu}$  as functions of the ${\rm S_{\rm e}}$, ${\rm S_{\rm e}}$ (in units of eV/$\AA$) for pure H$_2$O and CO ice mantles at CPS and CAPS, respectively. $\alpha$ = 1 and $\alpha$ = 0.6 are  reduction parameters on ${\rm S_{\rm e}}$ values for $\delta$ electrons energy loss process. $\rm a_{\rm gef,1}$, $\rm a_{\rm gef,4}$, and $\rm a_{\rm gef,10}$ are effective grain radii for first, fourth, and  tenth grain populations at the end of CPS and CAPS. Lines (blue circle: $\rm a_{\rm gef,1}$,red square: $\rm a_{\rm gef,4}$, green triangle: $\rm a_{\rm gef,10}$). The colored regions show the standard error estimations on yields. }
\label{fig:fig6_quad_yields_with_sem}
\end{figure*}
We confirm a strong correlation between ${\rm R_{\rm max,track}}$ and the $\gamma$ reduction factor. Hence, in calculating ${\rm R_{\rm max,track}}$, the $\gamma$ factor-dependent CR ion velocity effect should be included. This is due to the fact that the gradual decline in the $\gamma$ factor, scaled with $\rm \varepsilon(\rm Mev/nuc)$, directly determines the conversion efficiency of the deposited electronic energy as thermal energy. The variations of $\gamma$ factor may lead to a different amount of ${\rm R_{\rm max,track}}$ reduction even for the same electronic stopping power. When comparing to the same incident CR-type, we find that the effect of the $\gamma$ reduction factor on ${\rm R_{\rm max,track}}$ is more prominent for CO ice mantle with respect to H$_2$O because the latent track formation within CO ice is allowed for the broader CR kinetic energy range. 

\subsection{The Sputtering Efficiency}
\label{sec:SE}
According to the ATS model \citep{szenes1997amorphous,szenes2011comparison}, the homogeneous and cylindrical track is not allowed below a certain ${\rm S_{\rm e}}$ threshold (${\rm S_{\rm e,thres}}$). The recent experiment results achieved by D18 and D21 confirm the ice sputtering proceeds quadratically as a function of ${\rm S_{\rm e}}$, and the quadratic sputtering is closely related to cylindrical latent track formation. Besides, the experimental study of \citet{toulemonde2004track} shows that the latent track morphology can evolve from the extended spherical sub-components to the single-piece homogeneous cylindrical depending on the ${\rm S_{\rm e}}$ and on the physicochemical properties of the irradiated material. Furthermore, several molecular dynamic calculations \citep[e,g.,][]{bringa1999molecular,johnson1991linear,beuve2003influence} agree that low excitation energy densities within a CR ion track induced by low ${\rm S_{\rm e}}$ values may lead to linear or sub-linear sputtering.  Even though   the definition of an explicit transition between the linear and the quadratic sputtering regimes is challenging, as mentioned by \citet{dartois2020electronic}, linking up all these outcomes, we assume that the cylindrical latent track with continuous geometry is necessary for quadratic sputtering, whereas in the linear regime, the latent track consists of discontinuous spherical local components across the path of the incident CR ion. Therefore, we use the rudimentary argument that the homogeneous cylindrical latent track formation within ice mantles is a separator for the transition of the quadratic and the linear sputtering regimes. 
\begin{table*}
\caption{Total hot spot sputtering rate coefficients for H$_2$O and CO. }              
\centering                   
\begin{tabular}{c| c c c c c |c c c c c}      
\hline\hline            
\multicolumn{10}{c}{\textbf{EPS}, ice mantle type: H$_2$O, $\alpha = 0.6$ }\\ 
\multicolumn{5}{c}{${\rm k_{\rm hs}}$,(total) $\times 10^{-16} \;  [\rm molecules / s]$ $^{~a}$}&\multicolumn{6}{c}{$\rm f_{\rm hs}(\rm qu/li)$ $^{~b}$ }\\
\hline
$a_{gef}\;[\mu m]$ & $\rm k_{hs,W93_{200}}$  & $\rm k_{hs,W93_{400}}$  & $\rm k_{hs,W93_{600}}$ & $\rm k_{hs,P18_{low}}$  & $\rm k_{hs,P18_{high}}$ & $\rm f_{200}$  & $\rm f_{400}$  & $\rm f_{600}$ & $\rm f_{low}$   & $\rm f_{high}$  \\                      
\hline
0.039 &	3.528  & 0.487 & 0.151  & 1.069 & 203.051  & 0.370 & 0.330 & 0.313 & 0.416 & 0.447 \\ 
0.047 &	2.527  & 0.349 & 0.108  & 0.765 & 146.686  & 0.318 & 0.284 & 0.270 & 0.356 & 0.377 \\ 
0.057 &	2.203  & 0.305 & 0.094  & 0.666 & 129.775  & 0.276 & 0.247 & 0.235 & 0.308 & 0.319 \\ 
0.070 &	2.043  & 0.283 & 0.088  & 0.618 & 121.726  & 0.243 & 0.218 & 0.208 & 0.271 & 0.276 \\ 
0.087 &	1.947  & 0.270 & 0.084  & 0.588 & 116.461  & 0.220 & 0.197 & 0.187 & 0.245 & 0.248 \\ 
0.109 &	1.877  & 0.261 & 0.081  & 0.566 & 112.540  & 0.201 & 0.180 & 0.171 & 0.224 & 0.226 \\ 
0.136 &	1.819  & 0.253 & 0.079  & 0.548 & 109.535  & 0.184 & 0.165 & 0.157 & 0.205 & 0.205 \\ 
0.170 &	1.803  & 0.252 & 0.078  & 0.543 & 108.367  & 0.175 & 0.157 & 0.149 & 0.195 & 0.196 \\ 
0.213 &	1.792  & 0.251 & 0.078  & 0.539 & 107.406  & 0.167 & 0.149 & 0.141 & 0.186 & 0.187 \\ 
0.267 &	1.782  & 0.250 & 0.078  & 0.535 & 106.532  & 0.159 & 0.141 & 0.133 & 0.177 & 0.178 \\ 
\hline\hline  
\multicolumn{10}{c}{\textbf{CPS}, ice mantle type: H$_2$O, $\alpha = 0.6$ }\\ 
\multicolumn{5}{c}{${\rm k_{\rm hs}}$,(total) $\times 10^{-16} \;  [\rm molecules / s]$ }&\multicolumn{6}{c}{$\rm f_{\rm hs}(\rm qu/li)$ }\\
\hline
$a_{gef}\;[\mu m]$ & $\rm k_{hs,W93_{200}}$  & $\rm k_{hs,W93_{400}}$  & $\rm k_{hs,W93_{600}}$ & $\rm k_{hs,P18_{low}}$  & $\rm k_{hs,P18_{high}}$ & $\rm f_{200}$  & $\rm f_{400}$  & $\rm f_{600}$ & $\rm f_{low}$   & $\rm f_{high}$  \\ 
\hline
0.066 & 5.488 & 0.738 & 0.226 & 1.705 & 330.883 & 0.834	& 0.749	& 0.714	& 0.934	& 0.993 \\ 
0.071 & 3.962 & 0.533 & 0.163 & 1.230 & 240.517 & 0.765	& 0.689	& 0.658	& 0.855	& 0.894 \\ 
0.079 & 3.455 & 0.465 & 0.142 & 1.072 & 212.088 & 0.703	& 0.635	& 0.607	& 0.782	& 0.800 \\ 
0.09  & 3.212 & 0.433 & 0.132 & 0.997 & 198.978 & 0.65	& 0.589	& 0.564	& 0.722	& 0.725 \\ 
0.105 & 3.087 & 0.416 & 0.127 & 0.957 & 191.895 & 0.611	& 0.555	& 0.531	& 0.678	& 0.675 \\ 
0.124 & 3.011 & 0.406 & 0.124 & 0.932 & 187.448 & 0.578	& 0.525	& 0.502	& 0.640	& 0.633 \\ 
0.15  & 2.958 & 0.399 & 0.122 & 0.915 & 184.64  & 0.544	& 0.494	& 0.473	& 0.601	& 0.592 \\ 
0.182 & 2.931 & 0.396 & 0.121 & 0.906 & 183.57  & 0.513	& 0.466	& 0.447	& 0.566	& 0.554 \\ 
0.224 & 2.914 & 0.394 & 0.121 & 0.900 & 183.037 & 0.482	& 0.439	& 0.42	& 0.532	& 0.518 \\ 
0.277 & 2.925 & 0.396 & 0.121 & 0.903 & 183.873 & 0.455	& 0.414	& 0.397	& 0.502	& 0.487 \\ 
\hline\hline
\multicolumn{10}{c}{\textbf{CAPS}, ice mantle type: CO, $\alpha = 0.6$ }\\ 
\multicolumn{5}{c}{${\rm k_{\rm hs}}$,(total) $\times 10^{-16} \;  [\rm molecules / s]$}&\multicolumn{6}{c}{$\rm f_{\rm hs}(\rm qu/li)$}\\
\hline
$a_{gef}\;[\mu m]$ & $\rm k_{hs,W93_{200}}$  & $\rm k_{hs,W93_{400}}$  & $\rm k_{hs,W93_{600}}$ & $\rm k_{hs,P18_{low}}$  & $\rm k_{hs,P18_{high}}$ & $\rm f_{200}$  & $\rm f_{400}$  & $\rm f_{600}$ & $\rm f_{low}$   & $\rm f_{high}$  \\ 
\hline
0.078 & 82.979 & 10.812	& 3.252	& 25.293 & 5157.085	& 37.096 & 31.497 &	29.437 & 43.308 & 159.937 \\  
0.082 & 57.39  & 7.476	& 2.248	& 17.540 & 3634.349	& 34.313 & 29.302 &	27.448 & 40.131 & 146.084 \\  
0.089 & 48.02  & 6.255	& 1.881	& 14.702 & 3082.214	& 31.829 & 27.284 &	25.594 & 37.254 & 133.502 \\  
0.099 & 43.112 & 5.616	& 1.689	& 13.213 & 2791.228	& 29.675 & 25.497 &	23.94  & 34.761 & 123.673 \\  
0.112 & 39.171 & 5.105	& 1.536	& 12.010 & 2551.952	& 27.662 & 23.798 &	22.356 & 32.421 & 115.292 \\  
0.131 & 36.149 & 4.713	& 1.418	& 11.082 & 2357.758	& 26.051 & 22.429 &	21.075 & 30.527 & 108.096 \\  
0.156 & 33.90  & 4.423	& 1.331	& 10.384 & 2205.468	& 24.663 & 21.233 &	19.951 & 28.89	& 102.068 \\  
0.188 & 32.420 & 4.232	& 1.274	& 9.921	 & 2101.588	& 23.491 & 20.221 &	18.998 & 27.501 & 97.193  \\  
0.229 & 30.893 & 4.036	& 1.215	& 9.45	 & 2002.624	& 21.543 & 18.629 &	17.532 & 25.18	& 88.730  \\  
0.282 & 29.509 & 3.857	& 1.162	& 9.018	 & 1907.153	& 20.553 & 17.772 &	16.725 & 24.006 & 84.253  \\  
\hline   
\end{tabular}
\label{tab:table3_khs_results}  \\
\begin{itemize}[left=0\parindent]
\item[a :] The grain size-dependent total sputtering rate coefficients of H$_2$O and CO at three ice formation states as the summing of the sputtering components in the quadratic and the linear regimes. ${\rm k_{\rm hs}}$,(total) values in units of $10^{-16} \;  [\rm molecules / s]$ are derived from five CR spectra with different low-energy contents named as $\rm W93_{200}$,$\rm W93_{400}$, $\rm W93_{600}$,$\rm P18_{\rm low}$ and $\rm P18_{\rm high}$, respectively.
\item[b :] $\rm f_{\rm hs}(\rm qu/li)$ corresponds to the ratio of quadratic and linear sputtering rate coefficients for each CR energy distribution function and the grain size bin.
\end{itemize}
\flushleft\textit{Note:} The discrepancy between the highest and lowest value of ${\rm k_{\rm hs}}$ is due to the number of low-energy CR ions (see in text).
\end{table*} 

Figure \ref{fig:fig5_quad_yiled_variation} shows the quadratic sputtering yield variation as a function of CR ion kinetic energy (in units of MeV/nuc). To represent the characteristic relation between  the quadratic sputtering yield, $\rm Y_{\rm hs}^{\rm qu}$ and $\rm \varepsilon_{\rm ion}$, we give icy grain examples with the effective radii of 0.093 and 0.099 $\mu$m for CPS and CAPS, respectively. The variations in the quadratic sputtering yields are mainly driven by the ice binding energy and CR ion type and ${\rm S_{\rm e,eff}}$. 

 As can be seen in the figure, the quadratic H$_2$O sputtering yields are much lower than CO values. The relatively light ions lead to an efficient sputtering for CO ice at CAPS, whereas H$_2$O ice sputtering at CPS is induced by only CR ions with $\rm Z_{i} \geq 14$. CO ice sputtering can be easily produced within the broader $\rm \varepsilon$ range. $\rm Y_{\rm hs}^{\rm qu}$ values of H$_2$O and CO ice molecules increase proportionally to the atomic mass of the CR ion. In addition, when the $\alpha$ factor is reduced to 0.6 (the panels on the right), the quadratic yields significantly decrease for both CPS and CAPS.

The escaping of swift $\delta$ electrons from the icy surface may lead to the reduction of ${\rm S_{\rm e}}$ values in the range 0\% - 50\% (see \citet{johnson2013sputtering} and \citet{bringa2004new} for details). According to $\delta$ electron energy losses, the $\alpha$ factor can vary between 1 and 0.5. From the L85 results in appendix B, we exclude the dependency of the $\alpha$ factor with respect to the CR ion type, the CR kinetic energy range, and grain sizes. However, to examine the effect of delta electron losses on the sputtering yields, we choose 0.6 and 1 values of the $\alpha$ factor as the minimum and the maximum limits.

The ability of electron-phonon coupling is related to the $\gamma$ factor, which depends on the CR ion velocity. For this reason, the $\gamma$ factor directly controls the conversion efficiency between deposited electronic energy and thermal energy (see \citet{tombrello1994predicting} for details). In the quadratic sputtering regime, the $\gamma$ factor is also necessary for the calculation of the maximum radius of the latent track. In accordance with \citet{wesch2016ion} and \citet{szenes1997amorphous}, we suggest that the variation of $\gamma$ with the CR ion kinetic energy per nucleon ($\rm \varepsilon$) has three separated behaviors: for $\rm \varepsilon$ < 2 MeV/nuc, the $\gamma$ (the maximum value) is fixed and equal to 0.4, and for $\rm \varepsilon$  > 10 MeV/nuc (the minimum value) the $\gamma$ is still fixed but equal to 0.17, and the $\gamma$ shows a smooth decline for the intermediate values between 2 and 10 MeV/nuc. 
To represent the smooth reduction of  the $\gamma$ for the intermediate values, we use individually linear regression approximations over the whole kinetic energy range (1 MeV - 10 GeV) for each of 30 CR ions.

We determine that the sputtering yields can be enhanced by at least a factor of two when the CR ion-grain collision occurs at a non-normal incident angle.  This is because the effective mantle surface area seen by the incident CR ion increases with the increase of the CR ion impact angle with respect to the grain surface normal. Hence, the $\zeta$ increment factor should be included in the sputtering yield calculations. We suggest that the cylindrical latent track formation plays a critical role in the $\zeta$ variation.  As expected, the alteration characteristic of the $\zeta$ is different in the quadratic and the linear sputtering regimes. We also find that the $\zeta$ factor varies depending on the mantle composition, the grain size, and the ice mantle formation state. However, the influences of the last two parameters on the $\zeta$ are not dominant as the mantle composition.

Figure \ref{fig:fig6_quad_yields_with_sem} shows the quadratic sputtering, $\rm Y_{\rm hs}^{\rm qu}$ as a function of ${\rm S_{\rm e}}$. For all four panels in Figure \ref{fig:fig6_quad_yields_with_sem}, the calculated $\rm Y_{\rm hs}^{\rm qu}$ values are plotted with respect to ${\rm S_{\rm e}}$ ranges between the first minimums and the peak points where the lowest and the maximum quadratic yields are produced for the incident Fe CR ion. We find the size-dependent $\rm Y_{\rm hs}^{\rm qu}$ - ${\rm S_{\rm e}}$ relations are also similar for the other considered CR ion types relevant to the quadratic sputtering of H$_2$O and CO ices. As seen in Figure \ref{fig:fig6_quad_yields_with_sem}, even though the incident CR ion type (Fe) is the same, $\rm Y_{\rm hs}^{\rm qu}$ values exhibit apparent diversities that are directly linked to both the ice binding energy and the variation of ${\rm S_{\rm e}}$. 

We confirm a certain correlation between the sputtering efficiency and the grain size-dependent mantle depth evolution. For both H$_2$O and CO ice mantles, the sputtering yields tend to decrease with increasing grain size. This decrease originates from exponential decay and scales with a ratio of ${\rm D_{\rm ice,max}}$ to ${\rm D_{\rm p,ice}}$. We define this as a factor n($\chi$), see also section \ref{sec:HSIS}. The maximum ice mantle depth increment, which is inversely proportional to the grain size, can enhance the sputtering efficiency even for the same characteristic probe depth. 

We find that the effect of the n($\chi$) factor on yields is more apparent for CO with respect to H$_2$O because derived ${\rm D_{\rm p,ice}}$ values for CO are not only considerably higher at the same ${\rm S_{\rm e,eff}}$ but also $\rm D_{p,ice}(CO)$ values are constrained by two criteria. These criteria are the adopted CO mantle fraction and the size-dependent mantle formation. However, the effects of size dependency on the sputtering rate coefficient for both mantle types are relatively small compared to the variation of the CR ion spectrum.

CR ion-grain collision timescales are inversely proportional to grain cross-section (see Appendix \ref{sec:appC}). Thus, bigger grains are struck more frequently by CR ions. It is worth noting that despite the size dependency of collision frequency, sputtering rate coefficients derived from Equation (\ref{eq:Equation 15}) are independent of the grain cross-section. In evaluating sputtering yields, we consider the ratio of the effective grain surface area to the number of surface binding sites (the ${\rm q_{\rm MRN}}$) rather than the grain cross-section.
\begin{figure*}
\resizebox{0.95\textwidth}{!}{
\centering
\includegraphics[scale=0.48]{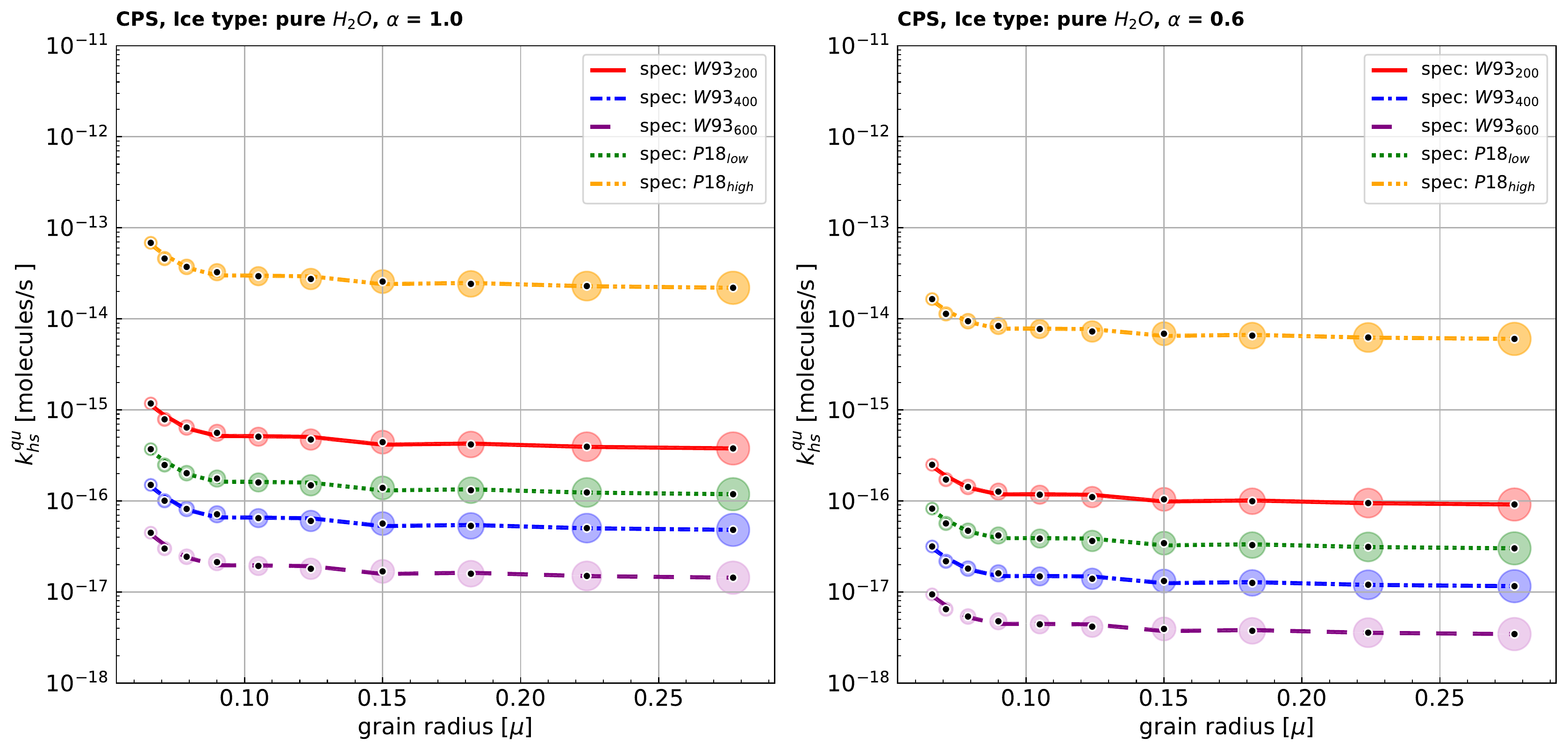}}
\caption{The quadratic cumulative sputtering rate coefficients, $\rm k_{\rm hs}^{\rm qu}$ (in units of molecules/s) of H$_2$O ice mantle for ten grain size bins and five CR ion energy spectra at CPS. $\alpha$ = 1 and $\alpha$ = 0.6 are  reduction parameters on ${\rm S_{\rm e}}$ values for $\delta$ electrons energy loss process. Black circles are the CR ion spectrum dependent  $\rm k_{\rm hs}^{\rm qu}$ values that are individually calculated  for ten grain effective radii in the range of 0.070 – 0.279 $\mu m$. The colored circular regions show the ice mantle evolution of the grain, scaled with the initial effective MRN radii before the ice formation. Five lines are the polynomial fits that are obtained to represent the triple relation between  $\rm k_{\rm hs}^{\rm qu}$ and the CR ion flux and the grain size.}
\label{fig:fig7_hs_rate_coeff_for_CPS}
\end{figure*}
\begin{figure*}
\resizebox{0.95\textwidth}{!}{
\centering
\includegraphics[scale=0.48]{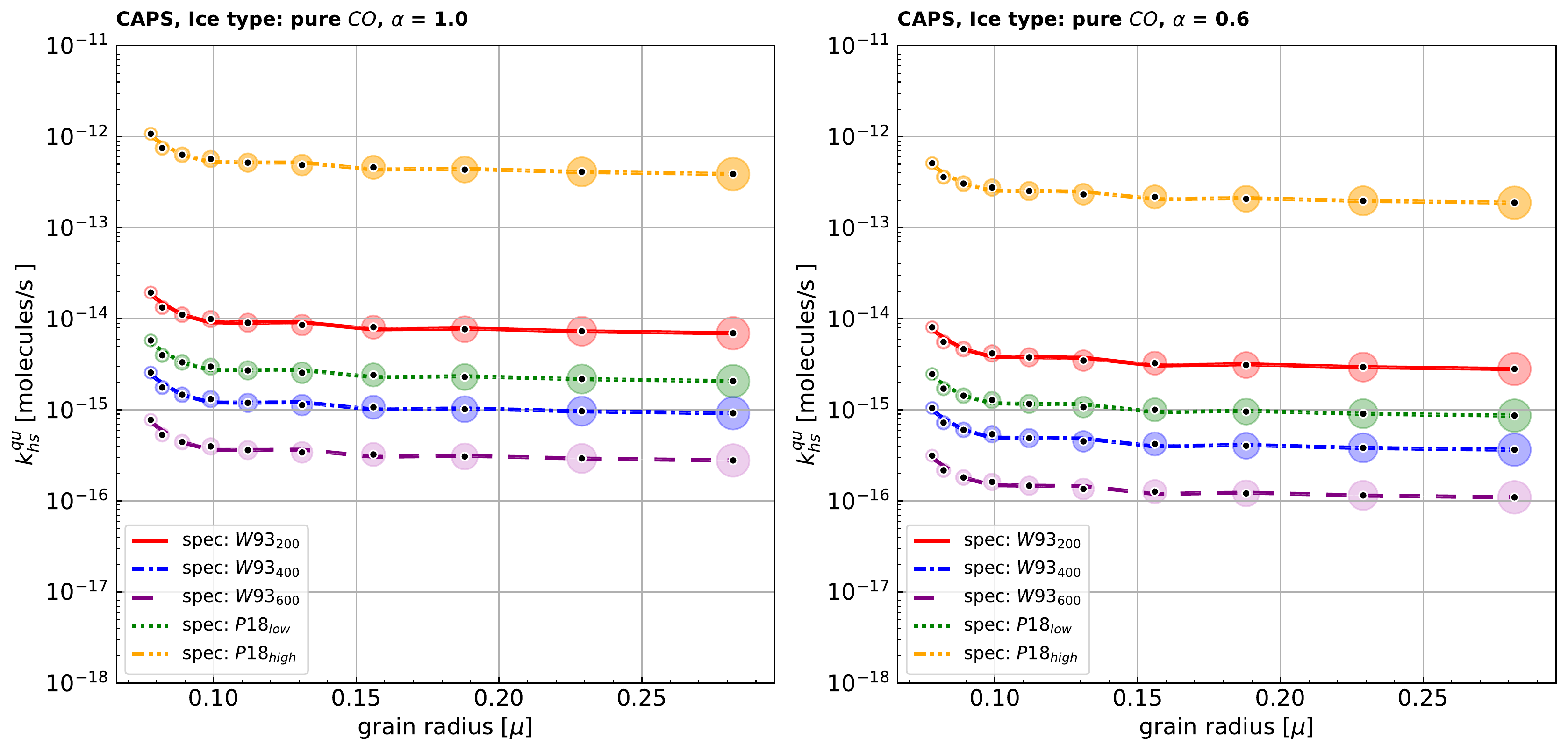}}
\caption{The quadratic cumulative sputtering rate coefficients,  $\rm k_{\rm hs}^{\rm qu}$ (in units of molecules/s) of CO ice mantle for ten grain size bins and five CR ion energy spectra at CAPS. The figure configuration is the same as in Figure \ref{fig:fig7_hs_rate_coeff_for_CPS}, except that the considered effective grain sizes replace with CAPS radii in the range of 0.078 – 0.282 $\mu m$. As in H$_2$O  ice, the calculated $\rm k_{\rm hs}^{\rm qu}$ values of CO ice dramatically depend on  the CR ion flux. However, in all conditions, the quadratic sputtering is more efficient for CO ice, roughly ten times with respect to H$_2$O ice.}
\label{fig:fig8_hs_rate_coeff_for_CAPS}
\end{figure*}

To verify our sputtering yields for CO ice, we compare our results as shown in Figures \ref{fig:fig5_quad_yiled_variation} and \ref{fig:fig6_quad_yields_with_sem} (bottom panels) with the experimental study of \citet[see their figures 6 and 7]{duarte2010laboratory}. One can see that our CO sputtering profiles as a function of the incident CR ion kinetic energy (Figure \ref{fig:fig5_quad_yiled_variation}), which are derived in our work, and our CO sputtering yield behavior as a function of $S_{\rm e}$ (Figure \ref{fig:fig6_quad_yields_with_sem}), exhibit clear similarities with respect to the results of \citet{duarte2010laboratory}. One of our model predictions is that the quadratic regime dominates CO sputtering in the case of a heavy CR ion - CO ice interaction.  We underline that this prediction is strengthened by absolute values of sputtering yields given in Figure \ref{fig:fig6_quad_yields_with_sem}, which are  analogous to the results of \citet[][see their figure 6]{duarte2010laboratory}.

It is worth noting that there are also some minor differences between our results and that of \citet{duarte2010laboratory}, especially the shape of the sputtering yield profile. There are several possible explanations for these differences, such as our choice of the n($\chi$)-dependent yield function (see Equation \ref{eq:Equation 13}), the effects of reduction factors $\alpha$ and $\gamma$ on $S_e$, and the adopted values for $\rm E_{\rm bind, ice}$. In summary, the main difference is attributed to our sputtering yield criteria of cylindrical latent track formation transitioning from a linear to a quadratic regime.

Table \ref{tab:table3_khs_results} shows the total sputtering rate coefficients, ${\rm k_{\rm hs}}$,(total), derived from five CR energy distribution functions for H$_2$O and CO at three ice formation states and in the case of $\alpha = 0.6$. We argue that the sputtering characteristics of H$_2$O and CO ice mantles are different because of two reasons. First,  CO sputtering efficiency is high as expected due to its low binding energy. Second, the quadratic regime mainly controls CO sputtering rate coefficients, while the linear regime gives nonnegligible contributions to the sputtering rate coefficients for H$_2$O depending on the CR ion spectrum, $\alpha$ reduction parameters, and the size-dependent ice formation state. 

When taking into account $\rm W93_{200}$ (at EPS) and $\rm W93_{400}$ (at CPS and CAPS) spectra, the calculated hot spot induced sputtering rate coefficients ($\rm k_{\rm hs}$) for H$_2$O and CO ice are consistent, at least on the order of magnitude with the results of \citet{bringa2004new} and \citet{silsbee2021ice}.  However, since there are distinct differences in the calculation methodology between our work and theirs, a direct comparison may not be meaningful.

Since we assume the cylindrical latent track formation as a separator transition between the quadratic and the linear sputtering regimes, the quadratic CO sputtering overcomes the linear sputtering. This is due to the fact that the cylindrical latent track formation within CO is possible even for lower ${\rm S_{\rm e,eff}}$ values, light CR ions, and the broader kinetic energy ranges. However, the cylindrical latent track formation for H$_2$O  is constrained by only heavy CR ions and higher ${\rm S_{\rm e,eff}}$ values within the narrower kinetic energy ranges. Therefore, the collective contributions of light CR ions and lower ${\rm S_{\rm e,eff}}$ values that are unsuitable for the continuous latent track formation enhance the efficiency of linear H$_2$O sputtering. 

As seen in Figures. \ref{fig:fig7_hs_rate_coeff_for_CPS} and \ref{fig:fig8_hs_rate_coeff_for_CAPS}, the sputtering efficiency is primarily governed by differential fluxes of CR ions, in other words, by CR spectra. The cumulative CR spectrum properties are formed mainly by two quantities: the abundance dependency of the CR spectrum and the effect of low-energy CR ions on the spectrum.

In the quadratic sputtering regime, a strong relation exists between the hot spot rate coefficient (${\rm k_{\rm hs}}$) and proton number of the CR ion ($\rm Z_{i}$) as $\rm k_{\rm hs} \sim  \rm Z_{i}^4$. This is because ${\rm S_{\rm e,eff}}$ values evolve quadratically as a function $\rm Z_{i}$ for H$_2$O and CO ice mantles. In light of this, we expect that heavy CR ions are more suitable for effective sputtering. However, this strong correlation is partially balanced by the lower abundance of heavy CR ions, in the range of two to five orders of magnitude, with respect to proton CR. 

To consider the variation of the low-energy part in the spectrum, which primarily manages hot spot-induced sputtering efficiency, we use five CR spectra with different low-energy CR ion contents. These spectra are named $\rm W93_{\rm 200}$, $\rm W93_{\rm 400}$, $\rm W93_{\rm 600}$, $\rm P18_{\rm low}$, and $\rm P18_{\rm high}$, respectively (see Figure \ref{fig:fig3_CR_spectra}).
$\rm P18_{\rm high}$ spectrum exhibits a more steep gradient at low kinetic energies contrary to other spectra, which means the weightiness of low-energy CR ions is the highest in $\rm P18_{\rm high}$, whereas $\rm P18_{\rm low}$ spectrum has a lower level of low-energy content with respect to $\rm P18_{\rm high}$ due to its smooth gradient at the low-energy part.

For W93 spectra, the low energy CR ion flux is controlled by a single parameter, defined as $\rm E_{C}$. Decrement of $\rm E_{C}$ leads to a significant increase in the efficiency of low-energy content in the spectrum. 
However, alterations in $\rm E_{C}$ have an almost negligible effect on the high-energy part of the spectrum and bigger $\rm E_{C}$ values correspond to less low-energy CR ions.  We use three $\rm E_{\rm C}$ of 200, 400, and 600 MeV, which roughly equates to the high ($\approx 3 \times 10^{-16} \; \rm s^{-1} $), the medium ($\approx  6 \times 10^{-17} \; \rm s^{-1}$), and the low ($\approx  2 \times 10^{-17} \; \rm s^{-1}$) ionization rates, respectively. 

The highest low-energy content spectrum, $\rm P18_{\rm high}$ results in very high grain size-dependent sputtering rate coefficients, scaled with $\cong \times 10^{-13}$ molecules/s for CO. These high values (which may not be realistic) are enough to completely desorb CO ice mantles into the gas phase in timescales of $\approx 10^{-2}$ million years for all adopted grain size bins. The lowest low-energy content spectrum, $\rm W93_{600}$ results in very low grain size-dependent sputtering rate coefficients, scaled with $\cong \times 10^{-16}$ molecules/s for even CO in timescales of $\approx 10^{2} -10^{3} $ million years. That means in the case of $\rm W93_{600}$, the hot spot-induced sputtering is too weak for efficient desorption of ice molecules in a typical molecular cloud lifetime. 

\section{CONCLUSIONS}
\label{sec:CONC}
In this study, we investigated the cosmic-ray induced sputtering process on icy grains in dense molecular cloud cores. We examined the formation conditions of the hot spot induced cylindrical latent track region within the ice mantles of grains. We also considered the grain size and the cosmic-ray flux dependencies on the sputtering efficiencies.

In our calculation routine, we first determined the threshold values of electronic stopping power $S_{\rm e, thres}$ required for the formation of cylindrical latent tracks. We then calculated the sputtering rate coefficients for the ice mantles around olivine grains, thereby also considering various correction factors that can play a role in the sputtering efficiencies. In our calculation space, there are 8 different parameters resulting in 58 dimensions, totalling to 81\,k combinations. The parameters are:
\begin{itemize}[left=0\parindent]
\item [-]Two pure ice mantle compositions (H$_2$O and CO).
\item [-]Three ice mantle formation states (H$_{2}$O-
dominated polar state in the edge: EPS, H2O-dominated polar state
in the center: CPS, and CO-dominated apolar state in the center:
CAPS).
\item [-]Ten grain sizes, ranging from 0.03$\mu {\rm m}$ to 0.3$\mu {\rm m}$.
\item [-]Thirty CR ion types (H to Zn), with specific abundances.
\item [-]Two cases for $\alpha$ reduction factor ($\alpha_{\rm max}$ = 1.0 and $\alpha_{\rm min}$= 0.6), where $\alpha$ corresponds to the energy loss due to induced secondary electrons.
\item [-]Three cases for CR ion kinetic energy dependent $\gamma$ reduction factor, where $\gamma$ designates the conversion efficiency of the deposited electronic energy to thermal energy.
\item [-]Three cases for CR ion incident angle-dependent average sputtering yield increment, $\overline{\zeta}$.
\item [-]{Five different CR ion energy spectra.}
\end{itemize}

Using these conditions, we obtained our sputtering yields (shown in Figures \ref{fig:fig5_quad_yiled_variation} and \ref{fig:fig6_quad_yields_with_sem}) and sputtering rate coefficients (shown in Figures. \ref{fig:fig7_hs_rate_coeff_for_CPS} and \ref{fig:fig8_hs_rate_coeff_for_CAPS}). A summary of results is as follows

\begin{itemize}[left=0\parindent]
\item [\textbf{1.}] The sputtering is at least ten times more efficient for CO with respect to that of H$_2$O, because the binding energy of CO is very low as compared to H$_2$O. 
For both H$_2$O and CO ice mantles, the sputtering efficiency is notably sensitive to the variation of effective electronic stopping power, ${\rm S_{\rm e,eff}}$ and the differential CR ion flux.
\item [\textbf{2.}] The sputtering efficiency is governed by CR ion properties (the atomic mass, the abundance, and the kinetic energy range) as well as the ice mantle composition, as expected. We calculate that in the case of $\alpha$ =1, the quadratic sputtering rate coefficients, $\rm k_{\rm hs}^{\rm qu}$(CO) at CAPS are on average 30 and 10 times higher than  values of $\rm k_{\rm hs}^{\rm qu}$(H$_2$O) at EPS and CPS, respectively. Whereas in the case of $\alpha$ = 0.6, the difference between $\rm k_{\rm hs}^{\rm qu}$(CO) and $\rm k_{\rm hs}^{\rm qu}$(H$_2$O) values are increased by almost a factor of 1.7 with respect to $\alpha$ =1 case.
\item [\textbf{3.}] The effect of the exponential decay factor n($\chi$) on the sputtering efficiency is more explicit for CO ice because of three reasons. First, the n($\chi$)  factor varies as a function of the ${\rm D_{\rm ice,max}}$/ ${\rm D_{\rm p,ice}}$ ratio, where $\rm D_{\rm ice, max}$ and $\rm D_{\rm p, ice}$ are the maximum ice mantle depth and the characteristic sputtering probe depth, respectively. Second, the obtained ${\rm D_{\rm p,ice}}$ values for CO ice are much higher than H$_2$O values. Third, ${\rm D_{\rm ice,max}}$ on the individual grain with effective radius $\rm a_{\rm grep}$ is restricted by the scaled division of total ice abundance into the grain size bins according to the MRN distribution at the three ice formation states.
\item [\textbf{4.}] An indirect correlation exists between the hot spot induced sputtering rate coefficient (${\rm k_{\rm hs}}$) and the grain sizes, derived from the MRN distribution. However, this correlation does not lead to dramatic alterations in ${\rm k_{\rm hs}}$ for the specific ten-grain sizes at the same ice mantle formation state because the grain size dependency of ${\rm k_{\rm hs}}$ completely arises from the exponential decay factor, n($\chi$), rather than the direct effect of grain size distribution. Therefore, we argued that for both H$_2$O and CO ice mantles, the grain size-dependent mantle depth evolution at three ice formation states gives minor contributions to the sputtering efficiency.
\item [\textbf{5.}] The characteristic sputtering behaviors of H$_2$O  and CO ice mantles are quite different. We suggested that CO sputtering is mainly controlled by the quadratic regime, whereas the sputtering contributions that come from the quadratic and the linear regimes are competitive for H$_2$O. This difference results from the adopted transition criterion between the quadratic and the linear sputtering regimes.
\item [\textbf{6.}] The track formation within CO ice can be produced by even light CR ions and the CO track formation is allowed for the broader ${\rm S_{\rm e,eff}}$ ranges. Therefore, the quadratic regime dominates CO sputtering. However, the quadratic sputtering and the track formation within H$_2$O ice are restricted by only heavy CR ions with the narrower ${\rm S_{\rm e,eff}}$ ranges. Thus, the cumulative contributions that originated from the light CR ions and the low ${\rm S_{\rm e,eff}}$ values below the track formation threshold increase the linear sputtering efficiency for H$_2$O depending on the differential CR ion flux, $\alpha$ reduction parameter, and the grain size.
\item [\textbf{7.}]  
Selecting a proper CR spectrum is necessary to avoid overestimating the CR spectrum-dependent hot spot sputtering efficiency for H$_2$O and CO ice mantles. For example, $\rm P18_{\rm high}$ spectrum results in immense coefficient values which may not be realistic. However, since the lower-energy part of the CR spectrum can be notably affected by attenuation processes, which we did not consider in this work, the different CR energy spectra are possible depending on the environmental conditions. Therefore, to ensure consistency with the adopted cloud core conditions, we suggested that $\rm W93_{200}$ at EPS and $\rm W93_{400}$ at CPS and CAPS are appropriate CR spectra for the qualifying of H$_2$O  and CO sputtering characteristics.
\end{itemize}
For our results, piecing together the arguments from  J91 and \citet{toulemonde2004track}, we inferred that the cylindrical latent track with continuous morphology is essential for quadratic sputtering, while in the linear sputtering regime, the latent track consists of localized and discontinuous spherical components. Adopting this somewhat simplified approximation, we suggest that the continuous cylindrical latent track formation within H$_2$O and CO ice mantles, which we identified by ${\rm S_{\rm e,thres}}$, can be used as a separator for the transition between the quadratic and the linear sputtering regimes. 

\section*{Acknowledgements}
ÖA thanks Mustafa Kürşad Yıldız and Cenk Kayhan for the editorial revisions, Maria Elisabetta Palumbo for help in calculation stopping power of ice mixtures, Kedron Silsbee for his constructive criticisms on the results of the sputtering yields, and Olli Sipilä for his suggestions about the definitions of the environmental conditions in our model.
\section*{DATA AVAILABILITY}
The data underlying this paper will be shared on reasonable request to the corresponding author.



\bibliographystyle{mnras}
\bibliography{References} 

\begin{thebibliography}{}
\makeatletter
\relax
\def\mn@urlcharsother{\let\do\@makeother \do\$\do\&\do\#\do\^\do\_\do\%\do\~}
\def\mn@doi{\begingroup\mn@urlcharsother \@ifnextchar [ {\mn@doi@}
  {\mn@doi@[]}}
\def\mn@doi@[#1]#2{\def\@tempa{#1}\ifx\@tempa\@empty \href
  {http://dx.doi.org/#2} {doi:#2}\else \href {http://dx.doi.org/#2} {#1}\fi
  \endgroup}
\def\mn@eprint#1#2{\mn@eprint@#1:#2::\@nil}
\def\mn@eprint@arXiv#1{\href {http://arxiv.org/abs/#1} {{\tt arXiv:#1}}}
\def\mn@eprint@dblp#1{\href {http://dblp.uni-trier.de/rec/bibtex/#1.xml}
  {dblp:#1}}
\def\mn@eprint@#1:#2:#3:#4\@nil{\def\@tempa {#1}\def\@tempb {#2}\def\@tempc
  {#3}\ifx \@tempc \@empty \let \@tempc \@tempb \let \@tempb \@tempa \fi \ifx
  \@tempb \@empty \def\@tempb {arXiv}\fi \@ifundefined
  {mn@eprint@\@tempb}{\@tempb:\@tempc}{\expandafter \expandafter \csname
  mn@eprint@\@tempb\endcsname \expandafter{\@tempc}}}

\bibitem[\protect\citeauthoryear{Abplanalp, F{\"o}rstel  \& Kaiser}{Abplanalp
  et~al.}{2016}]{abplanalp2016exploiting}
Abplanalp M.~J.,  F{\"o}rstel M.,   Kaiser R.~I.,  2016, Chemical Physics
  Letters, 644, 79

\bibitem[\protect\citeauthoryear{Agull{\'o}-L{\'o}pez, Garc{\'\i}a  \&
  Olivares}{Agull{\'o}-L{\'o}pez et~al.}{2005}]{agullo2005lattice}
Agull{\'o}-L{\'o}pez F.,  Garc{\'\i}a G.,   Olivares J.,  2005, Journal of
  applied physics, 97, 093514

\bibitem[\protect\citeauthoryear{Anders \& Urbassek}{Anders \&
  Urbassek}{2013}]{anders2013impacts}
Anders C.,  Urbassek H.~M.,  2013, Nuclear Instruments and Methods in Physics
  Research Section B: Beam Interactions with Materials and Atoms, 303, 200

\bibitem[\protect\citeauthoryear{Anders \& Urbassek}{Anders \&
  Urbassek}{2019a}]{anders2019energetic}
Anders C.,  Urbassek H.~M.,  2019a, Monthly Notices of the Royal Astronomical
  Society, 482, 2374

\bibitem[\protect\citeauthoryear{Anders \& Urbassek}{Anders \&
  Urbassek}{2019b}]{anders2019high}
Anders C.,  Urbassek H.~M.,  2019b, Astronomy \& Astrophysics, 625, A140

\bibitem[\protect\citeauthoryear{Anders, Bringa  \& Urbassek}{Anders
  et~al.}{2020}]{anders2020ejection}
Anders C.,  Bringa E.~M.,   Urbassek H.~M.,  2020, The Astrophysical Journal,
  891, 21

\bibitem[\protect\citeauthoryear{Andersson, Al-Halabi, Kroes  \& van
  Dishoeck}{Andersson et~al.}{2006}]{andersson2006molecular}
Andersson S.,  Al-Halabi A.,  Kroes G.-J.,   van Dishoeck E.~F.,  2006, The
  Journal of chemical physics, 124, 064715

\bibitem[\protect\citeauthoryear{Baragiola, Vidal, Svendsen, Schou, Shi, Bahr
  \& Atteberrry}{Baragiola et~al.}{2003}]{baragiola2003sputtering}
Baragiola R.~A.,  Vidal R.~A.,  Svendsen W.,  Schou J.,  Shi M.,  Bahr D.,
  Atteberrry C.,  2003, Nuclear Instruments and Methods in Physics Research
  Section B: Beam Interactions with Materials and Atoms, 209, 294

\bibitem[\protect\citeauthoryear{Beuve, Stolterfoht, Toulemonde, Trautmann  \&
  Urbassek}{Beuve et~al.}{2003}]{beuve2003influence}
Beuve M.,  Stolterfoht N.,  Toulemonde M.,  Trautmann C.,   Urbassek H.~M.,
  2003, Physical Review B, 68, 125423

\bibitem[\protect\citeauthoryear{Binns et~al.,}{Binns
  et~al.}{2014}]{binns2014supertiger}
Binns W.,  et~al., 2014, The Astrophysical Journal, 788, 18

\bibitem[\protect\citeauthoryear{Boogert et~al.,}{Boogert
  et~al.}{2011}]{boogert2011ice}
Boogert A.,  et~al., 2011, The Astrophysical Journal, 729, 92

\bibitem[\protect\citeauthoryear{Boogert, Gerakines  \& Whittet}{Boogert
  et~al.}{2015}]{boogert2015observations}
Boogert A.~A.,  Gerakines P.~A.,   Whittet D.~C.,  2015, Annual Review of
  Astronomy and Astrophysics, 53

\bibitem[\protect\citeauthoryear{Bringa \& Johnson}{Bringa \&
  Johnson}{2001}]{bringa2001angular}
Bringa E.,  Johnson R.,  2001, Nuclear Instruments and Methods in Physics
  Research Section B: Beam Interactions with Materials and Atoms, 180, 99

\bibitem[\protect\citeauthoryear{Bringa \& Johnson}{Bringa \&
  Johnson}{2002}]{bringa2002sputtering}
Bringa E.,  Johnson R.,  2002, Nuclear Instruments and Methods in Physics
  Research Section B: Beam Interactions with Materials and Atoms, 193, 365

\bibitem[\protect\citeauthoryear{Bringa \& Johnson}{Bringa \&
  Johnson}{2004}]{bringa2004new}
Bringa E.~M.,  Johnson R.~E.,  2004, The Astrophysical Journal, 603, 159

\bibitem[\protect\citeauthoryear{Bringa, Johnson  et~al.}{Bringa
  et~al.}{1999}]{bringa1999molecular}
Bringa E.,  Johnson R.,   et~al., 1999, Nuclear Instruments and Methods in
  Physics Research Section B: Beam Interactions with Materials and Atoms, 152,
  267

\bibitem[\protect\citeauthoryear{Bringa et~al.,}{Bringa
  et~al.}{2007}]{bringa2007energetic}
Bringa E.,  et~al., 2007, The Astrophysical Journal, 662, 372

\bibitem[\protect\citeauthoryear{Brown \& Charnley}{Brown \&
  Charnley}{1990}]{brown1990chemical}
Brown P.~D.,  Charnley S.,  1990, Monthly Notices of the Royal Astronomical
  Society, 244, 432

\bibitem[\protect\citeauthoryear{Brown, Lanzerotti, Poate  \&
  Augustyniak}{Brown et~al.}{1978}]{brown1978sputtering}
Brown W.,  Lanzerotti L.,  Poate J.,   Augustyniak W.,  1978, Physical Review
  Letters, 40, 1027

\bibitem[\protect\citeauthoryear{Brown, Augustyniak, Lanzerotti, Johnson  \&
  Evatt}{Brown et~al.}{1980}]{brown1980linear}
Brown W.,  Augustyniak W.,  Lanzerotti L.,  Johnson R.,   Evatt R.,  1980,
  Physical Review Letters, 45, 1632

\bibitem[\protect\citeauthoryear{Caselli, Walmsley, Tafalla, Dore  \&
  Myers}{Caselli et~al.}{1999}]{caselli1999co}
Caselli P.,  Walmsley C.,  Tafalla M.,  Dore L.,   Myers P.,  1999, The
  Astrophysical Journal, 523, L165

\bibitem[\protect\citeauthoryear{Caselli et~al.,}{Caselli
  et~al.}{2012}]{caselli2012first}
Caselli P.,  et~al., 2012, The Astrophysical journal letters, 759, L37

\bibitem[\protect\citeauthoryear{Caselli et~al.,}{Caselli
  et~al.}{2022}]{caselli2022central}
Caselli P.,  et~al., 2022, The Astrophysical Journal, 929, 13

\bibitem[\protect\citeauthoryear{Cazaux, Minissale, Dulieu  \& Hocuk}{Cazaux
  et~al.}{2016}]{cazaux2016dust}
Cazaux S.,  Minissale M.,  Dulieu F.,   Hocuk S.,  2016, Astronomy \&
  Astrophysics, 585, A55

\bibitem[\protect\citeauthoryear{Chabot}{Chabot}{2016}]{chabot2016cosmic}
Chabot M.,  2016, Astronomy \& Astrophysics, 585, A15

\bibitem[\protect\citeauthoryear{Chac{\'o}n-Tanarro et~al.,}{Chac{\'o}n-Tanarro
  et~al.}{2019}]{chacon2019dust}
Chac{\'o}n-Tanarro A.,  et~al., 2019, Astronomy \& Astrophysics, 623, A118

\bibitem[\protect\citeauthoryear{Compi{\`e}gne et~al.,}{Compi{\`e}gne
  et~al.}{2011}]{compiegne2011global}
Compi{\`e}gne M.,  et~al., 2011, Astronomy \& Astrophysics, 525, A103

\bibitem[\protect\citeauthoryear{Cuppen, Penteado, Isokoski, van~der Marel  \&
  Linnartz}{Cuppen et~al.}{2011}]{cuppen2011co}
Cuppen H.,  Penteado E.,  Isokoski K.,  van~der Marel N.,   Linnartz H.,  2011,
  Monthly Notices of the Royal Astronomical Society, 417, 2809

\bibitem[\protect\citeauthoryear{Cuppen, Walsh, Lamberts, Semenov, Garrod,
  Penteado  \& Ioppolo}{Cuppen et~al.}{2017}]{cuppen2017grain}
Cuppen H.,  Walsh C.,  Lamberts T.,  Semenov D.,  Garrod R.,  Penteado E.,
  Ioppolo S.,  2017, Space Science Reviews, 212, 1

\bibitem[\protect\citeauthoryear{Dartois et~al.,}{Dartois
  et~al.}{2013}]{dartois2013swift}
Dartois E.,  et~al., 2013, Astronomy \& Astrophysics, 557, A97

\bibitem[\protect\citeauthoryear{Dartois et~al.,}{Dartois
  et~al.}{2015}]{dartois2015heavy}
Dartois E.,  et~al., 2015, Astronomy \& Astrophysics, 576, A125

\bibitem[\protect\citeauthoryear{Dartois, Chabot, Barkach, Rothard, Aug{\'e},
  Agnihotri, Domaracka  \& Boduch}{Dartois et~al.}{2018}]{dartois2018cosmic}
Dartois E.,  Chabot M.,  Barkach T.~I.,  Rothard H.,  Aug{\'e} B.,  Agnihotri
  A.,  Domaracka A.,   Boduch P.,  2018, Astronomy \& Astrophysics, 618, A173

\bibitem[\protect\citeauthoryear{Dartois, Chabot, Barkach, Rothard, Aug{\'e},
  Agnihotri, Domaracka  \& Boduch}{Dartois et~al.}{2019}]{dartois2019non}
Dartois E.,  Chabot M.,  Barkach T.~I.,  Rothard H.,  Aug{\'e} B.,  Agnihotri
  A.,  Domaracka A.,   Boduch P.,  2019, Astronomy \& Astrophysics, 627, A55

\bibitem[\protect\citeauthoryear{Dartois, Chabot, Barkach, Rothard, Boduch,
  Aug{\'e}, Duprat  \& Rojas}{Dartois et~al.}{2020}]{dartois2020electronic}
Dartois E.,  Chabot M.,  Barkach T.~I.,  Rothard H.,  Boduch P.,  Aug{\'e} B.,
  Duprat J.,   Rojas J.,  2020, Nuclear Instruments and Methods in Physics
  Research Section B: Beam Interactions with Materials and Atoms, 485, 13

\bibitem[\protect\citeauthoryear{Dartois, Chabot, Barkach, Rothard, Boduch,
  Aug{\'e}  \& Agnihotri}{Dartois et~al.}{2021}]{dartois2021cosmic}
Dartois E.,  Chabot M.,  Barkach T.~I.,  Rothard H.,  Boduch P.,  Aug{\'e} B.,
   Agnihotri A.,  2021, Astronomy \& Astrophysics, 647, A177

\bibitem[\protect\citeauthoryear{De~Jong \& Kamijo}{De~Jong \&
  Kamijo}{1973}]{de1973growth}
De~Jong T.,  Kamijo F.,  1973, Astronomy and Astrophysics, 25

\bibitem[\protect\citeauthoryear{Draine}{Draine}{2010}]{draine2010physics}
Draine B.~T.,  2010, Physics of the interstellar and intergalactic medium.
Princeton University Press

\bibitem[\protect\citeauthoryear{Draine \& Li}{Draine \&
  Li}{2007}]{draine2007infrared}
Draine B.,  Li A.,  2007, The Astrophysical Journal, 657, 810

\bibitem[\protect\citeauthoryear{Duarte, Domaracka, Boduch, Rothard, Dartois
  \& Da~Silveira}{Duarte et~al.}{2010}]{duarte2010laboratory}
Duarte E.~S.,  Domaracka A.,  Boduch P.,  Rothard H.,  Dartois E.,
  Da~Silveira E.,  2010, Astronomy \& Astrophysics, 512, A71

\bibitem[\protect\citeauthoryear{Ehrenfreund et~al.,}{Ehrenfreund
  et~al.}{1999}]{ehrenfreund1999laboratory}
Ehrenfreund P.,  et~al., 1999, Astronomy and Astrophysics, 350, 240

\bibitem[\protect\citeauthoryear{Erents \& McCracken}{Erents \&
  McCracken}{1973}]{erents1973desorption}
Erents S.,  McCracken G.,  1973, Journal of Applied Physics, 44, 3139

\bibitem[\protect\citeauthoryear{Fraser, Collings, Dever  \& McCoustra}{Fraser
  et~al.}{2004}]{fraser2004using}
Fraser H.~J.,  Collings M.~P.,  Dever J.~W.,   McCoustra M.~R.,  2004, Monthly
  Notices of the Royal Astronomical Society, 353, 59

\bibitem[\protect\citeauthoryear{Gabici, Evoli, Gaggero, Lipari, Mertsch,
  Orlando, Strong  \& Vittino}{Gabici et~al.}{2019}]{gabici2019origin}
Gabici S.,  Evoli C.,  Gaggero D.,  Lipari P.,  Mertsch P.,  Orlando E.,
  Strong A.,   Vittino A.,  2019, arXiv preprint arXiv:1903.11584

\bibitem[\protect\citeauthoryear{Garrod}{Garrod}{2008}]{garrod2008new}
Garrod R.,  2008, Astronomy \& Astrophysics, 491, 239

\bibitem[\protect\citeauthoryear{Garrod \& Herbst}{Garrod \&
  Herbst}{2006}]{garrod2006formation}
Garrod R.,  Herbst E.,  2006, Astronomy \& Astrophysics, 457, 927

\bibitem[\protect\citeauthoryear{Gorai et~al.,}{Gorai
  et~al.}{2020}]{gorai2020systematic}
Gorai P.,  et~al., 2020, ACS Earth and Space Chemistry

\bibitem[\protect\citeauthoryear{Gupta, Ganesan, Sulania  \& Das}{Gupta
  et~al.}{2017}]{gupta2017role}
Gupta S.,  Ganesan V.,  Sulania I.,   Das B.,  2017, Surface Science, 664, 137

\bibitem[\protect\citeauthoryear{G{\"u}ver \& {\"O}zel}{G{\"u}ver \&
  {\"O}zel}{2009}]{guver2009relation}
G{\"u}ver T.,  {\"O}zel F.,  2009, Monthly Notices of the Royal Astronomical
  Society, 400, 2050

\bibitem[\protect\citeauthoryear{Habing}{Habing}{1968}]{habing1968interstellar}
Habing H.,  1968, Bulletin of the Astronomical Institutes of the Netherlands,
  19, 421

\bibitem[\protect\citeauthoryear{Hasegawa \& Herbst}{Hasegawa \&
  Herbst}{1993}]{hasegawa1993new}
Hasegawa T.~I.,  Herbst E.,  1993, Monthly Notices of the Royal Astronomical
  Society, 261, 83

\bibitem[\protect\citeauthoryear{Henning}{Henning}{2010}]{henning2010cosmic}
Henning T.,  2010, Annual Review of Astronomy and Astrophysics, 48, 21

\bibitem[\protect\citeauthoryear{Herbst, Chang  \& Cuppen}{Herbst
  et~al.}{2005}]{herbst2005chemistry}
Herbst E.,  Chang Q.,   Cuppen H.,  2005, in Journal of Physics Conference
  Series. pp 18--35

\bibitem[\protect\citeauthoryear{Hocuk \& Cazaux}{Hocuk \&
  Cazaux}{2015}]{hocuk2015interplay}
Hocuk S.,  Cazaux S.,  2015, Astronomy \& Astrophysics, 576, A49

\bibitem[\protect\citeauthoryear{Hocuk, Cazaux, Spaans  \& Caselli}{Hocuk
  et~al.}{2016}]{hocuk2016chemistry}
Hocuk S.,  Cazaux S.,  Spaans M.,   Caselli P.,  2016, Monthly Notices of the
  Royal Astronomical Society, 456, 2586

\bibitem[\protect\citeauthoryear{Hocuk, Sz{\H{u}}cs, Caselli, Cazaux, Spaans
  \& Esplugues}{Hocuk et~al.}{2017}]{hocuk2017parameterizing}
Hocuk S.,  Sz{\H{u}}cs L.,  Caselli P.,  Cazaux S.,  Spaans M.,   Esplugues G.,
   2017, Astronomy \& Astrophysics, 604, A58

\bibitem[\protect\citeauthoryear{Hollenbach, Kaufman, Bergin  \&
  Melnick}{Hollenbach et~al.}{2008}]{hollenbach2008water}
Hollenbach D.,  Kaufman M.~J.,  Bergin E.~A.,   Melnick G.~J.,  2008, The
  Astrophysical Journal, 690, 1497

\bibitem[\protect\citeauthoryear{Ip \& Axford}{Ip \&
  Axford}{1985}]{ip1985estimates}
Ip W.-H.,  Axford W.,  1985, Astronomy and Astrophysics, 149, 7

\bibitem[\protect\citeauthoryear{Iqbal, Wakelam  \& Gratier}{Iqbal
  et~al.}{2018}]{iqbal2018statistical}
Iqbal W.,  Wakelam V.,   Gratier P.,  2018, Astronomy \& Astrophysics, 620,
  A109

\bibitem[\protect\citeauthoryear{Ivlev, R{\"o}cker, Vasyunin  \& Caselli}{Ivlev
  et~al.}{2015}]{ivlev2015impulsive}
Ivlev A.,  R{\"o}cker T.,  Vasyunin A.,   Caselli P.,  2015, The Astrophysical
  Journal, 805, 59

\bibitem[\protect\citeauthoryear{Johnson, Pospieszalska  \& Brown}{Johnson
  et~al.}{1991}]{johnson1991linear}
Johnson R.,  Pospieszalska M.,   Brown W.,  1991, Physical Review B, 44, 7263

\bibitem[\protect\citeauthoryear{Johnson, Carlson, Cassidy  \& Fama}{Johnson
  et~al.}{2013}]{johnson2013sputtering}
Johnson R.~E.,  Carlson R.~W.,  Cassidy T.~A.,   Fama M.,  2013, in , The
  science of solar system ices.
Springer, pp 551--581

\bibitem[\protect\citeauthoryear{Jones \& Williams}{Jones \&
  Williams}{1984}]{jones19843}
Jones A.,  Williams D.,  1984, Monthly Notices of the Royal Astronomical
  Society, 209, 955

\bibitem[\protect\citeauthoryear{J{\o}rgensen, Sch{\"o}ier  \&
  Van~Dishoeck}{J{\o}rgensen et~al.}{2005}]{jorgensen2005molecular}
J{\o}rgensen J.,  Sch{\"o}ier F.,   Van~Dishoeck E.,  2005, Astronomy \&
  Astrophysics, 435, 177

\bibitem[\protect\citeauthoryear{Jurac, Johnson  \& Donn}{Jurac
  et~al.}{1998}]{jurac1998monte}
Jurac S.,  Johnson R.,   Donn B.,  1998, The Astrophysical Journal, 503, 247

\bibitem[\protect\citeauthoryear{Kalv{\=a}ns}{Kalv{\=a}ns}{2015a}]{kalvans2015cosmic}
Kalv{\=a}ns J.,  2015a, Astronomy \& Astrophysics, 573, A38

\bibitem[\protect\citeauthoryear{Kalv{\=a}ns}{Kalv{\=a}ns}{2015b}]{kalvans2015ice}
Kalv{\=a}ns J.,  2015b, The Astrophysical Journal, 806, 196

\bibitem[\protect\citeauthoryear{Kalv{\=a}ns}{Kalv{\=a}ns}{2018}]{kalvans2018temperature}
Kalv{\=a}ns J.,  2018, The Astrophysical Journal Supplement Series, 239, 6

\bibitem[\protect\citeauthoryear{Kalv{\=a}ns \& Kalnin}{Kalv{\=a}ns \&
  Kalnin}{2019}]{kalvans2019chemical}
Kalv{\=a}ns J.,  Kalnin J.~R.,  2019, Monthly Notices of the Royal Astronomical
  Society, 486, 2050

\bibitem[\protect\citeauthoryear{Kalv{\=a}ns \& Kalnin}{Kalv{\=a}ns \&
  Kalnin}{2020}]{kalvans2020evaporative}
Kalv{\=a}ns J.,  Kalnin J.~R.,  2020, Astronomy \& Astrophysics, 641, A49

\bibitem[\protect\citeauthoryear{K{\"o}hler, Ysard  \& Jones}{K{\"o}hler
  et~al.}{2015}]{kohler2015dust}
K{\"o}hler M.,  Ysard N.,   Jones A.~P.,  2015, Astronomy \& Astrophysics, 579,
  A15

\bibitem[\protect\citeauthoryear{Leger, Jura  \& Omont}{Leger
  et~al.}{1985}]{leger1985desorption}
Leger A.,  Jura M.,   Omont A.,  1985, Astronomy and Astrophysics, 144, 147

\bibitem[\protect\citeauthoryear{Li \& Draine}{Li \&
  Draine}{2001a}]{li2001ultrasmall}
Li A.,  Draine B.,  2001a, The Astrophysical Journal Letters, 550, L213

\bibitem[\protect\citeauthoryear{Li \& Draine}{Li \&
  Draine}{2001b}]{li2001infrared}
Li A.,  Draine B.,  2001b, The Astrophysical Journal, 554, 778

\bibitem[\protect\citeauthoryear{Lounis-Mokrani, Badreddine, Mebhah,
  Imatoukene, Fromm  \& Allab}{Lounis-Mokrani
  et~al.}{2008}]{lounis2008determination}
Lounis-Mokrani Z.,  Badreddine A.,  Mebhah D.,  Imatoukene D.,  Fromm M.,
  Allab M.,  2008, Radiation measurements, 43, S41

\bibitem[\protect\citeauthoryear{Mainitz, Anders  \& Urbassek}{Mainitz
  et~al.}{2016}]{mainitz2016irradiation}
Mainitz M.,  Anders C.,   Urbassek H.~M.,  2016, Astronomy \& Astrophysics,
  592, A35

\bibitem[\protect\citeauthoryear{Mainitz, Anders  \& Urbassek}{Mainitz
  et~al.}{2017}]{mainitz2017impact}
Mainitz M.,  Anders C.,   Urbassek H.~M.,  2017, Nuclear Instruments and
  Methods in Physics Research Section B: Beam Interactions with Materials and
  Atoms, 393, 34

\bibitem[\protect\citeauthoryear{Marsh et~al.,}{Marsh
  et~al.}{2014}]{marsh2014properties}
Marsh K.,  et~al., 2014, Monthly Notices of the Royal Astronomical Society,
  439, 3683

\bibitem[\protect\citeauthoryear{Mathis, Rumpl  \& Nordsieck}{Mathis
  et~al.}{1977}]{mathis1977size}
Mathis J.~S.,  Rumpl W.,   Nordsieck K.~H.,  1977, The Astrophysical Journal,
  217, 425

\bibitem[\protect\citeauthoryear{Meftah, Brisard, Costantini, Hage-Ali,
  Stoquert, Studer  \& Toulemonde}{Meftah et~al.}{1993}]{meftah1993swift}
Meftah A.,  Brisard F.,  Costantini J.,  Hage-Ali M.,  Stoquert J.,  Studer F.,
    Toulemonde M.,  1993, Physical Review B, 48, 920

\bibitem[\protect\citeauthoryear{Noble, Fraser, Pontoppidan  \& Craigon}{Noble
  et~al.}{2017}]{noble2017two}
Noble J.,  Fraser H.,  Pontoppidan K.,   Craigon A.,  2017, Monthly Notices of
  the Royal Astronomical Society, 467, 4753

\bibitem[\protect\citeauthoryear{{\"O}berg, Boogert, Pontoppidan, van~den
  Broek, van Dishoeck, Bottinelli, Blake  \& Evans}{{\"O}berg
  et~al.}{2011}]{oberg2011ices}
{\"O}berg K.~I.,  Boogert A.~A.,  Pontoppidan K.~M.,  van~den Broek S.,  van
  Dishoeck E.~F.,  Bottinelli S.,  Blake G.~A.,   Evans N.~J.,  2011,
  Proceedings of the International Astronomical Union, 7, 65

\bibitem[\protect\citeauthoryear{Ormel, Paszun, Dominik  \& Tielens}{Ormel
  et~al.}{2009}]{ormel2009dust}
Ormel C.,  Paszun D.,  Dominik C.,   Tielens A.,  2009, Astronomy \&
  Astrophysics, 502, 845

\bibitem[\protect\citeauthoryear{Ormel, Min, Tielens, Dominik  \& Paszun}{Ormel
  et~al.}{2011}]{ormel2011dust}
Ormel C.,  Min M.,  Tielens A.,  Dominik C.,   Paszun D.,  2011, Astronomy \&
  Astrophysics, 532, A43

\bibitem[\protect\citeauthoryear{Orsi, Collaboration  et~al.}{Orsi
  et~al.}{2007}]{orsi2007pamela}
Orsi S.,  Collaboration P.,   et~al., 2007, Nuclear Instruments and Methods in
  Physics Research Section A: Accelerators, Spectrometers, Detectors and
  Associated Equipment, 580, 880

\bibitem[\protect\citeauthoryear{Ossenkopf}{Ossenkopf}{1993}]{ossenkopf1993dust}
Ossenkopf V.,  1993, Astronomy and Astrophysics, 280, 617

\bibitem[\protect\citeauthoryear{Ossenkopf \& Henning}{Ossenkopf \&
  Henning}{1994}]{ossenkopf1994dust}
Ossenkopf V.,  Henning T.,  1994, Astronomy and Astrophysics, 291, 943

\bibitem[\protect\citeauthoryear{Padovani, Galli  \& Glassgold}{Padovani
  et~al.}{2009}]{padovani2009cosmic}
Padovani M.,  Galli D.,   Glassgold A.~E.,  2009, Astronomy \& Astrophysics,
  501, 619

\bibitem[\protect\citeauthoryear{Padovani, Ivlev, Galli  \& Caselli}{Padovani
  et~al.}{2018}]{padovani2018cosmic}
Padovani M.,  Ivlev A.~V.,  Galli D.,   Caselli P.,  2018, Astronomy \&
  Astrophysics, 614, A111

\bibitem[\protect\citeauthoryear{Palumbo \& Strazzulla}{Palumbo \&
  Strazzulla}{1993}]{palumbo19932140}
Palumbo M.,  Strazzulla G.,  1993, Astronomy and Astrophysics, 269, 568

\bibitem[\protect\citeauthoryear{Parikka, Juvela, Pelkonen, Malinen  \&
  Harju}{Parikka et~al.}{2015}]{parikka2015physical}
Parikka A.,  Juvela M.,  Pelkonen V.-M.,  Malinen J.,   Harju J.,  2015,
  Astronomy \& Astrophysics, 577, A69

\bibitem[\protect\citeauthoryear{Pauly \& Garrod}{Pauly \&
  Garrod}{2016}]{pauly2016effects}
Pauly T.,  Garrod R.~T.,  2016, The Astrophysical Journal, 817, 146

\bibitem[\protect\citeauthoryear{Pineda, Goldsmith, Chapman, Snell, Li,
  Cambr{\'e}sy  \& Brunt}{Pineda et~al.}{2010}]{pineda2010relation}
Pineda J.~L.,  Goldsmith P.~F.,  Chapman N.,  Snell R.~L.,  Li D.,
  Cambr{\'e}sy L.,   Brunt C.,  2010, The Astrophysical Journal, 721, 686

\bibitem[\protect\citeauthoryear{Pontoppidan}{Pontoppidan}{2006}]{pontoppidan2006spatial}
Pontoppidan K.,  2006, Astronomy \& Astrophysics, 453, L47

\bibitem[\protect\citeauthoryear{Pontoppidan et~al.,}{Pontoppidan
  et~al.}{2008}]{pontoppidan2008c2d}
Pontoppidan K.~M.,  et~al., 2008, The Astrophysical Journal, 678, 1005

\bibitem[\protect\citeauthoryear{Powers}{Powers}{1980}]{powers1980influence}
Powers D.,  1980, Accounts of Chemical Research, 13, 433

\bibitem[\protect\citeauthoryear{Qasim, Chuang, Fedoseev, Ioppolo, Boogert  \&
  Linnartz}{Qasim et~al.}{2018}]{qasim2018formation}
Qasim D.,  Chuang K.-J.,  Fedoseev G.,  Ioppolo S.,  Boogert A.,   Linnartz H.,
   2018, Astronomy \& Astrophysics, 612, A83

\bibitem[\protect\citeauthoryear{Reboussin, Wakelam, Guilloteau  \&
  Hersant}{Reboussin et~al.}{2014}]{reboussin2014grain}
Reboussin L.,  Wakelam V.,  Guilloteau S.,   Hersant F.,  2014, Monthly Notices
  of the Royal Astronomical Society, 440, 3557

\bibitem[\protect\citeauthoryear{Sabin \& Oddershede}{Sabin \&
  Oddershede}{2009}]{sabin2009analytical}
Sabin J.~R.,  Oddershede J.,  2009, International Journal of Quantum Chemistry,
  109, 2933

\bibitem[\protect\citeauthoryear{Schmalzl et~al.,}{Schmalzl
  et~al.}{2014}]{schmalzl2014earliest}
Schmalzl M.,  et~al., 2014, Astronomy \& Astrophysics, 569, A7

\bibitem[\protect\citeauthoryear{Schou, Bo, Ellegaard, So, Claussen
  et~al.}{Schou et~al.}{1986}]{schou1986erosion}
Schou J.,  Bo P.,  Ellegaard O.,  So H.,  Claussen C.,   et~al., 1986, Physical
  Review B, 34, 93

\bibitem[\protect\citeauthoryear{Schutte \& Greenberg}{Schutte \&
  Greenberg}{1991}]{schutte1991explosive}
Schutte W.,  Greenberg J.,  1991, Astronomy and Astrophysics, 244, 190

\bibitem[\protect\citeauthoryear{Shen, Greenberg, Schutte  \&
  Van~Dishoeck}{Shen et~al.}{2004}]{shen2004cosmic}
Shen C.,  Greenberg J.,  Schutte W.,   Van~Dishoeck E.,  2004, Astronomy \&
  Astrophysics, 415, 203

\bibitem[\protect\citeauthoryear{Shingledecker, Tennis, Le~Gal  \&
  Herbst}{Shingledecker et~al.}{2018}]{shingledecker2018cosmic}
Shingledecker C.~N.,  Tennis J.,  Le~Gal R.,   Herbst E.,  2018, The
  Astrophysical Journal, 861, 20

\bibitem[\protect\citeauthoryear{Shingledecker, Incerti, Ivlev, Emfietzoglou,
  Kyriakou, Vasyunin  \& Caselli}{Shingledecker
  et~al.}{2020}]{shingledecker2020cosmic}
Shingledecker C.~N.,  Incerti S.,  Ivlev A.,  Emfietzoglou D.,  Kyriakou I.,
  Vasyunin A.,   Caselli P.,  2020, The Astrophysical Journal, 904, 189

\bibitem[\protect\citeauthoryear{Sigmund}{Sigmund}{1969}]{sigmund1969theory}
Sigmund P.,  1969, Physical review, 184, 383

\bibitem[\protect\citeauthoryear{Sigmund}{Sigmund}{1987}]{sigmund1987mechanisms}
Sigmund P.,  1987, Nuclear Instruments and Methods in Physics Research Section
  B: Beam Interactions with Materials and Atoms, 27, 1

\bibitem[\protect\citeauthoryear{Silsbee, Caselli  \& Ivlev}{Silsbee
  et~al.}{2021}]{silsbee2021ice}
Silsbee K.,  Caselli P.,   Ivlev A.~V.,  2021, Monthly Notices of the Royal
  Astronomical Society, 507, 6205

\bibitem[\protect\citeauthoryear{Sipil{\"a}, Zhao  \& Caselli}{Sipil{\"a}
  et~al.}{2020}]{sipila2020effect}
Sipil{\"a} O.,  Zhao B.,   Caselli P.,  2020, Astronomy \& Astrophysics, 640,
  A94

\bibitem[\protect\citeauthoryear{Sipil{\"a}, Silsbee  \& Caselli}{Sipil{\"a}
  et~al.}{2021}]{sipila2021revised}
Sipil{\"a} O.,  Silsbee K.,   Caselli P.,  2021, The Astrophysical Journal,
  922, 126

\bibitem[\protect\citeauthoryear{Sofia \& Meyer}{Sofia \&
  Meyer}{2001}]{sofia2001erratum}
Sofia U.,  Meyer D.,  2001, The Astrophysical Journal, 558, L147

\bibitem[\protect\citeauthoryear{Steinacker et~al.,}{Steinacker
  et~al.}{2015}]{steinacker2015grain}
Steinacker J.,  et~al., 2015, Astronomy \& Astrophysics, 582, A70

\bibitem[\protect\citeauthoryear{Stone et~al.,}{Stone
  et~al.}{1998}]{stone1998cosmic}
Stone E.~C.,  et~al., 1998, in , The Advanced Composition Explorer Mission.
Springer, pp 285--356

\bibitem[\protect\citeauthoryear{Szenes}{Szenes}{1996}]{szenes1996formation}
Szenes G.,  1996, Nuclear Instruments and Methods in Physics Research Section
  B: Beam Interactions with Materials and Atoms, 107, 146

\bibitem[\protect\citeauthoryear{Szenes}{Szenes}{1997}]{szenes1997amorphous}
Szenes G.,  1997, Nuclear Instruments and Methods in Physics Research Section
  B: Beam Interactions with Materials and Atoms, 122, 530

\bibitem[\protect\citeauthoryear{Szenes}{Szenes}{2011}]{szenes2011comparison}
Szenes G.,  2011, Nuclear Instruments and Methods in Physics Research Section
  B: Beam Interactions with Materials and Atoms, 269, 174

\bibitem[\protect\citeauthoryear{Szenes, Horvath, Pecz, Paszti  \& Toth}{Szenes
  et~al.}{2002}]{szenes2002tracks}
Szenes G.,  Horvath Z.,  Pecz B.,  Paszti F.,   Toth L.,  2002, Physical Review
  B, 65, 045206

\bibitem[\protect\citeauthoryear{Takayanagi}{Takayanagi}{1973}]{takayanagi1973molecule}
Takayanagi K.,  1973, Publications of the Astronomical Society of Japan, 25,
  327

\bibitem[\protect\citeauthoryear{Tatischeff, Decourchelle  \&
  Maurin}{Tatischeff et~al.}{2012}]{tatischeff2012nonthermal}
Tatischeff V.,  Decourchelle A.,   Maurin G.,  2012, Astronomy \& Astrophysics,
  546, A88

\bibitem[\protect\citeauthoryear{Tielens}{Tielens}{2005}]{tielens2005physics}
Tielens A.~G.,  2005, The physics and chemistry of the interstellar medium.
Cambridge University Press

\bibitem[\protect\citeauthoryear{Tolstikhina, Imai, Winckler  \&
  Shevelko}{Tolstikhina et~al.}{2018}]{tolstikhina2018basic}
Tolstikhina I.,  Imai M.,  Winckler N.,   Shevelko V.,  2018, Springer Series
  on Atomic, Optical, and Plasma Physics, Springer

\bibitem[\protect\citeauthoryear{Tombrello}{Tombrello}{1994}]{tombrello1994predicting}
Tombrello T.,  1994, Nuclear Instruments and Methods in Physics Research
  Section B: Beam Interactions with Materials and Atoms, 94, 424

\bibitem[\protect\citeauthoryear{Toulemonde, Paumier  \& Dufour}{Toulemonde
  et~al.}{1993}]{toulemonde1993thermal}
Toulemonde M.,  Paumier E.,   Dufour C.,  1993, Radiation Effects and Defects
  in Solids, 126, 201

\bibitem[\protect\citeauthoryear{Toulemonde, Trautmann, Balanzat, Hjort  \&
  Weidinger}{Toulemonde et~al.}{2004}]{toulemonde2004track}
Toulemonde M.,  Trautmann C.,  Balanzat E.,  Hjort K.,   Weidinger A.,  2004,
  Nuclear Instruments and Methods in Physics Research Section B: Beam
  Interactions with Materials and Atoms, 216, 1

\bibitem[\protect\citeauthoryear{Valencic \& Smith}{Valencic \&
  Smith}{2015}]{valencic2015interstellar}
Valencic L.~A.,  Smith R.~K.,  2015, The Astrophysical Journal, 809, 66

\bibitem[\protect\citeauthoryear{Vasyunin \& Herbst}{Vasyunin \&
  Herbst}{2013}]{vasyunin2013reactive}
Vasyunin A.,  Herbst E.,  2013, The Astrophysical Journal, 769, 34

\bibitem[\protect\citeauthoryear{Wada, Tanaka, Suyama, Kimura  \&
  Yamamoto}{Wada et~al.}{2009}]{wada2009collisional}
Wada K.,  Tanaka H.,  Suyama T.,  Kimura H.,   Yamamoto T.,  2009, The
  Astrophysical Journal, 702, 1490

\bibitem[\protect\citeauthoryear{Wakelam, Dartois, Chabot, Spezzano,
  Navarro-Almaida, Loison  \& Fuente}{Wakelam
  et~al.}{2021}]{wakelam2021efficiency}
Wakelam V.,  Dartois E.,  Chabot M.,  Spezzano S.,  Navarro-Almaida D.,  Loison
  J.-C.,   Fuente A.,  2021, Astronomy \& Astrophysics, 652, A63

\bibitem[\protect\citeauthoryear{Walmsley \& Flower}{Walmsley \&
  Flower}{2004}]{walmsley2004complete}
Walmsley C.~M.,  Flower D.,  2004, Astronomy \& Astrophysics, 418, 1035

\bibitem[\protect\citeauthoryear{Watanabe, Nagaoka, Shiraki  \&
  Kouchi}{Watanabe et~al.}{2004}]{watanabe2004hydrogenation}
Watanabe N.,  Nagaoka A.,  Shiraki T.,   Kouchi A.,  2004, The Astrophysical
  Journal, 616, 638

\bibitem[\protect\citeauthoryear{Webber \& Yushak}{Webber \&
  Yushak}{1983}]{webber1983measurement}
Webber W.,  Yushak S.,  1983, The Astrophysical Journal, 275, 391

\bibitem[\protect\citeauthoryear{Weingartner \& Draine}{Weingartner \&
  Draine}{2001}]{weingartner2001dust}
Weingartner J.~C.,  Draine B.,  2001, The Astrophysical Journal, 548, 296

\bibitem[\protect\citeauthoryear{Wesch \& Wendler}{Wesch \&
  Wendler}{2016}]{wesch2016ion}
Wesch W.,  Wendler E.,  2016, Series in Surf. Sci, 61

\bibitem[\protect\citeauthoryear{Wettlaufer}{Wettlaufer}{2010}]{wettlaufer2010accretion}
Wettlaufer J.,  2010, The Astrophysical Journal, 719, 540

\bibitem[\protect\citeauthoryear{Wilms, Allen  \& McCray}{Wilms
  et~al.}{2000}]{wilms2000absorption}
Wilms J.,  Allen A.,   McCray R.,  2000, The Astrophysical Journal, 542, 914

\bibitem[\protect\citeauthoryear{Ysard, K{\"o}hler, Jones, Dartois, Godard  \&
  Gavilan}{Ysard et~al.}{2016}]{ysard2016mantle}
Ysard N.,  K{\"o}hler M.,  Jones A.,  Dartois E.,  Godard M.,   Gavilan L.,
  2016, Astronomy \& Astrophysics, 588, A44

\bibitem[\protect\citeauthoryear{Zhao, Caselli  \& Li}{Zhao
  et~al.}{2018}]{zhao2018effect}
Zhao B.,  Caselli P.,   Li Z.-Y.,  2018, Monthly Notices of the Royal
  Astronomical Society, 478, 2723

\bibitem[\protect\citeauthoryear{Zhu, Tian, Li  \& Zhang}{Zhu
  et~al.}{2017}]{zhu2017gas}
Zhu H.,  Tian W.,  Li A.,   Zhang M.,  2017, Monthly Notices of the Royal
  Astronomical Society, 471, 3494

\bibitem[\protect\citeauthoryear{Ziegler}{Ziegler}{2004}]{ziegler2004srim}
Ziegler J.~F.,  2004, Nuclear instruments and methods in physics research
  section B: Beam interactions with materials and atoms, 219, 1027

\bibitem[\protect\citeauthoryear{Ziegler, Ziegler  \& Biersack}{Ziegler
  et~al.}{2010}]{ziegler2010srim}
Ziegler J.~F.,  Ziegler M.~D.,   Biersack J.~P.,  2010, Nuclear Instruments and
  Methods in Physics Research Section B: Beam Interactions with Materials and
  Atoms, 268, 1818

\makeatother
\end{thebibliography}




\appendix
\section{Icy Grain Model Calculations}
\label{sec:appA}
Calculation routines for our icy grain model are given as follows:
\begin{itemize}[left=0\parindent]
\item [1.] \textit{The grain size distribution}:
When we consider the fixed dust-to-gas ratio $\rm R_{d}$, the total gas mass $\rm M_{gas}$ equals $\mu$ x $m_{p}$, where $\mu$ is the mean particle mass of the medium and $m_{p}$ $\approx$ $1.67 \times 10^{-24} \;g$ is the mass of the proton. Considering hydrogen is entirely found in molecular form and using the relative elemental compositions in table 2 of \citet{wilms2000absorption} for the medium, we assume that $\mu$ is equal to 2.35. 

The ${\rm C_{\rm MRN}}$ constant of the MRN distribution for the adopted grain size range  is calculated from
\begin{equation}\label{eq:A.1}
 \rm R_{d} M_{gas}       =  \rm C_{MRN}\int_{a_{gmin}}^{a_{gmax}}\rho_{g} \times V_{g} \times a_{g}^{-3.5} da_{g},
\end{equation}
\begin{equation}\label{eq:A.2}
\rm C_{MRN}              =  \rm \frac{R_{d} \,m_{p} \,\mu }{2 \, \pi \,\rho_{g}} \times \frac{3}{4} \times \Bigg[\frac{1}{\sqrt{a_{gmax}} - \sqrt{a_{gmin}}}\Bigg] \;\; \; \rm cm^{2.5}.
\end{equation}
\item [2.] \textit{The effective grain radius}:  We prefer the size intervals scaled with $\rm log \sigma_{g} = \rm log(\pi \times a_{g}^{2})$, because surface coverage of the ice mantle is proportional to the multiplication of the cross-sectional area of the substrate grain and the grain abundance \citep{cuppen2017grain}. Considering this size division, we calculate the grain abundance (with respect to hydrogen) and the grain cross-section in each size interval, namely the effective abundance and the effective cross-section  as follows:

The effective abundance, $X_{\rm d,k}$, in each size interval k is 
\begin{equation}\label{eq:A.3}
\rm X_{d,k} = \rm C_{MRN}\int_{a_{gmin,k}}^{a_{gmax,k}} a_{g}^{-3.5} da_{g}.
\end{equation}
The effective cross section, $\sigma_{\rm gef,k}$ in each size interval k is 
\begin{equation}\label{eq:A.4}
\rm \sigma_{gef,k} = \rm \frac{\int_{a_{gmin,k}}^{a_{gmax,k}} \pi \, a_{g}^{2} \times a_{g}^{-3.5} da_{g}}{X_{d,k} /C_{MRN}} \; \; \rm cm^{2}.
\end{equation}
Using Equation (\ref{eq:A.4}) for each size interval, we calculate the effective and initial radii of bare grains before ice mantle formation. The calculated initial radius values in units of $\rm \mu m$ for the ten size intervals are 0.033, 0.042, 0.053, 0.067, 0.084, 0.106, 0.133, 0.168, 0.211 and 0.266. Eventually, hierarchically summing up the size-dependent ice mantle depths and the initial bare grain radii, we individually derive the final grain radius as $\rm a_{\rm gef}$ in each ten size interval for different ice formation states. These results are given in Table \ref{tab:table2_MRN_grain_model}. 
\item [3.] \textit{The representative grain radius}: In obtaining the number of feasible binding sites on surfaces of grains in each size interval k (1 to 10), we use the multiplication of $ \mathrm{X_{d,k}\, 4\,\sigma_{gef,k}} $ and $\rm l_{s}^{-2}$.  We take a canonical value of $\rm l_{s} = 3 \, \AA$ \citep{herbst2005chemistry,cazaux2016dust,hocuk2016chemistry, pauly2016effects, zhao2018effect}.  $\rm X_{\rm d,rep}$, $\rm A_{\rm grep}$, and $\rm N_{\rm site,rep}$ are 
\begin{equation}\label{eq:A.5}
\rm X_{d,rep}     = \rm \frac{R_{d} M_{gas}}{\frac{4 \pi}{3}\, \rho_{g} \, a_{grep}^{3}},
\end{equation}
\begin{equation}\label{eq:A.6}
\rm A_{grep}      = \rm 4 \, \pi \, a_{grep}^{2} \; \; \rm cm^{2},
\end{equation}
\begin{equation}\label{eq:A.7}
\rm N_{site,grep}  = \rm X_{d,rep} \times A_{grep} \times l_{s}^{-2},
\end{equation}
where $X_{\rm d,rep}$ is the representative grain abundance, $\rm A_{\rm grep}$ is the representative grain surface area,  $\rm N_{\rm site,rep}$ is the representative number of surface binding sites on the grain.

When $\rm N_{\rm site,grep}$ is equalized to $\mathrm{\sum_{k =1,10}(X_{d,k}\, 4\,\sigma_{gef,k}) \,l_{s}^{-2}} $,
$a_{grep}$ is 
\begin{equation}\label{eq:A.8}
\rm a_{grep}     = \rm \frac{3 R_{d} M_{gas}}{\rho_{g}} \times \frac{1}{\sum_{k =1,10}(X_{d,k}\, 4\,\sigma_{gef,k})} \; \rm cm.
\end{equation}
\item [4.] \textit{The ice mantle depth}: We derive a value of $\rm D_{\rm ice,max,a_{grep}}$ $ \approx$ 289 $ \AA$. This specific ice mantle depth roughly corresponds to 96 ice monolayers around the grain with $\rm a_{\rm grep}$ radius when assuming that the physical depth of a single monolayer equals the $\rm l_{s}$. We then calculate the representative maximum ice abundance which coincides with $\rm n_{ice,max}(a_{grep})$ = $ 3.805 \times \, 10^{-4}$, using a relation is 
\begin{equation}\label{eq:A.9}
\rm n_{ice,max,a_{grep}} \,= \rm \, N_{mono,grep} \times\; N_{site,grep}.
\end{equation} 
$\rm N_{mono,grep} =  \rm D_{ice,max,a_{grep}}/ l_{s}$  is the number of maximum monolayers of 0.095 $\mu$m grain at the end of three ice formation states. 

The ice mantle formation-dependent grain size growth is inversely proportional to the grain radius. This is due to the fact that the total surface area of the MRN distribution is dominated by smaller grains \citep{ossenkopf1993dust,ormel2011dust,boogert2015observations, steinacker2015grain}. Considering this MRN distribution behavior, we adopt that the total maximum mantle depth reaches the greatest value for the smallest grain size interval. We calculate that the smallest grain size interval with a total number of surface binding sites, $\rm N_{site,gef,1}\,= \,6.276 \times \, 10^{-7}$ initially has a 0.033 $\mu$m effective radius. For that reason, we take  $\rm D_{ice,max,a_{gef,1}}$ = 450 $\AA$ as the total maximum depth limit of the smallest grain size interval at the end of three ice formation states, by solving Equation (\ref{eq:A.9}) according to the $\rm N_{site,gef,1}$ and a definite ice abundance value ($24.74\% \,\rm n_{ice,max,a_{grep}}$).
 
Consequently, we individually re-scale the total maximum mantle depths with respect to grain size intervals for three ice formation states using a scaling factor. The adopted scaling factor is 
\begin{equation}\label{eq:A.10}
\rm (D_{ice,max,a_{gef,1}})\times \; \Bigg[\frac{N_{site,gef,1}}{(X_{d,k}\, 4\,\sigma_{gef,k}) \,l_{s}^{-2})}\Bigg] \times \; \Bigg[\frac{d_{ice,max,j}}{d_{ice,max}}\Bigg],
\end{equation}
in each size bin k (1 to 10) and in each ice formation state j (a to c). According to this scaling, at the end of three ice formation states, the total radius increases nearly a factor of three (2.36) in the smallest effective size interval, whereas the total radius in the largest effective size interval is almost the same as the initial value.  The calculated ice mantle depths are given in Table \ref{tab:table2_MRN_grain_model}.
\end{itemize}
\section{The Impact Angle Correction}
\label{sec:appB}
In evaluating the CR ion impact angle-dependent sputtering yield increment factor, $\zeta$  as a function of $\rm [{cos \theta}]^{-x}$, where $\rm \theta$ is the impact angle between the incident direction of CR ion and the surface normal of target grain, we take three x values as 1, 1.6, and 1.7. B01 suggests that the power law index of $\zeta$ function is governed by the $\xi$ factor. Thus, $\zeta$ factor evolves depending on the ratio of the deposited radial energy density in the latent track to the surface binding energy density of the ice mantle. According to this assumption, x index equals  1, 1.6, and 1.7 for $\xi \geq 1$, $\xi\leq 1$ ,and $\xi \ll 1$ cases, respectively.  

As mentioned in Sections \ref{sec:LTF} and \ref{sec:SE}, the latent track radii within CO ice are much larger than H$_2$O values and CO sputtering is mainly controlled by the quadratic regime. Therefore, in the quadratic sputtering regime, $\zeta$ evolution is dominated  by $\xi\leq 1$  (x = 1.6) case for CO ice, whereas $\xi \geq 1$ (x = 1) case gives the main contribution to the zeta factor for H$_2$O ice. In the linear sputtering regime, we only take into account  $\xi \ll 1$ (x = 1.7) case for both H$_2$O and CO ice mantles. Because we expect that the radial energy density across the path of incident CR ion is quite lower with respect to the quadratic regime.

In considering the effect of grain size variation the  on $\zeta$ factor, we modify the equation of L85 in Appendix C for three $\xi$ cases. The analytical expression of average $\zeta$ factor over the impact solid angle range between 0 and ($\pi/2$ - z) is

\begin{equation}\label{eq:B.1}
       \mathrm{\overline{\zeta}} = \frac{\int_{0}^{\pi/2 -z} \;[{cos \theta}]^{-x} \times d\Omega}{\int_{0}^{\pi/2 -z}d\Omega}   =  \zeta \times (1-z)^{-x} \times ln\Big[\frac{1}{z^{-x}}\Big],
\,
   \end{equation}  
where z is a cut off value to provide the convergence of the integral result in each grain size bin k (1 to 10). In calculating z values, we assume that z is equal to $\mathrm{\sqrt[]{\frac{D_{ice}}{(a_{grep,k})}}}$.  ${\rm D_{\rm ice}}$ is $\rm D_{max,ice,k}$ for $\rm D_{p,ice} \geq D_{ice,max,k}$ and ${\rm D_{\rm ice}}$ is replaced to ${\rm D_{\rm p,ice}}$ when $\rm D_{p,ice} < D_{ice,max,k}$.

\begin{figure*}
\resizebox{0.96\textwidth}{!}{
\centering
\includegraphics[scale=0.48]{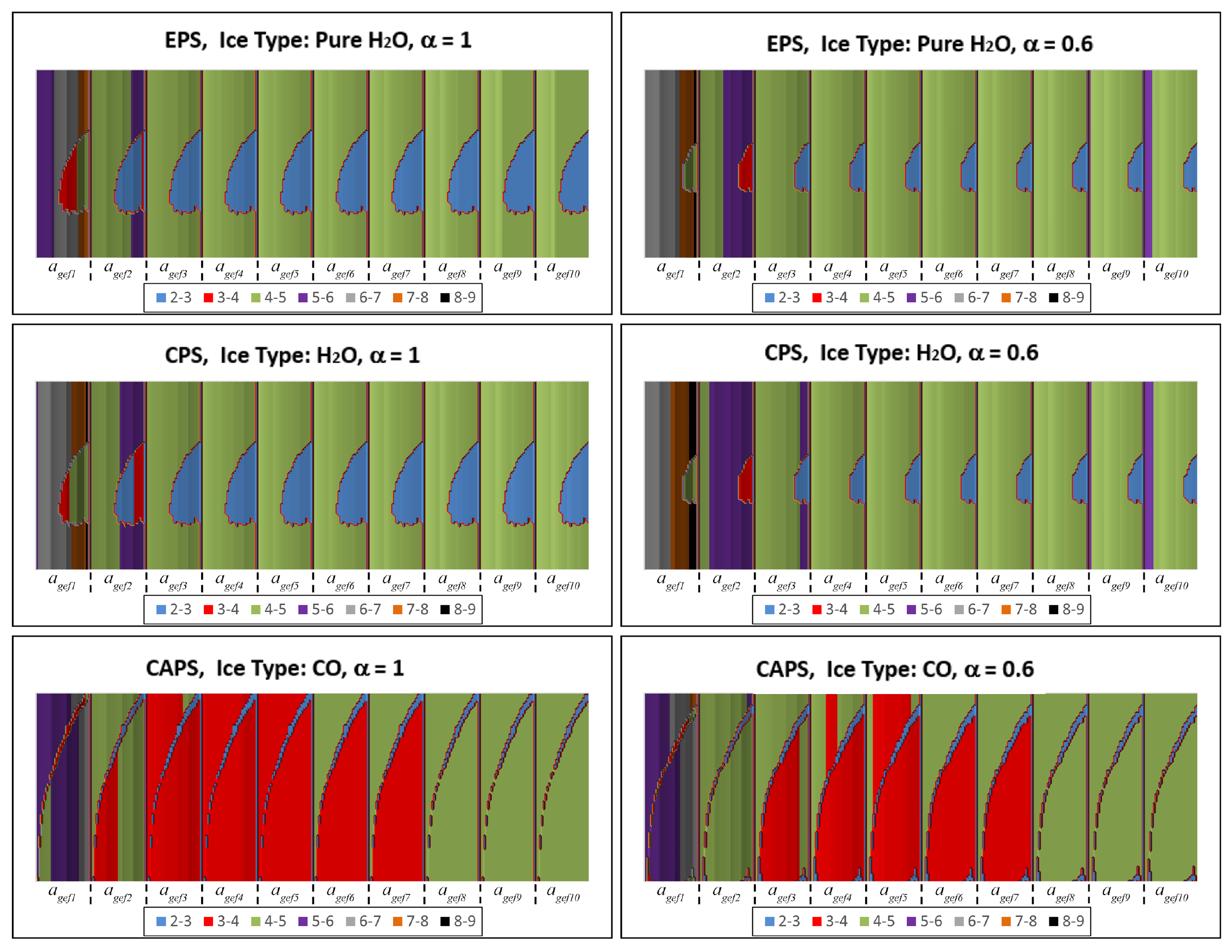}}
\caption{The evolution of average increment factor, $\rm \overline{\zeta}$, derived from three $\rm \xi$ cases for ten grain size bins at EPS, CPS and CAPS. }
\label{fig:fig9_zeta_evoluation}
\end{figure*}

Figure \ref{fig:fig9_zeta_evoluation} shows the calculated average grain size-dependent increment factors, $\overline{\zeta}$ for H$_2$O (at EPS and CPS) and CO (at CAPS) ices, according to three $\rm \xi$ cases.
\section{The Impact Time Scale}
\label{sec:appC}
\begin{figure*}
\begin{adjustbox}{addcode={\begin{minipage}{\width}}{\caption{The grain size and CR energy spectrum dependent impact time scales at EPS, CPS, and CAPS for five different CR ion energy spectra named as $\rm W93_{200}$, $\rm W93_{400}$, $\rm W93_{600}$, $\rm P18_{low}$ and $\rm P18_{high}$.  In EPS and CPS cases, we show the results of four CR ion types (Si, Cr, Fe, and Zn).  In the CAPS case, we show the results of five CR ion types (O, Si, Cr, Fe, and Zn ). CR impact times are scaled with million years for all samples.}\end{minipage}},rotate=90,center}
\includegraphics[scale=.80]{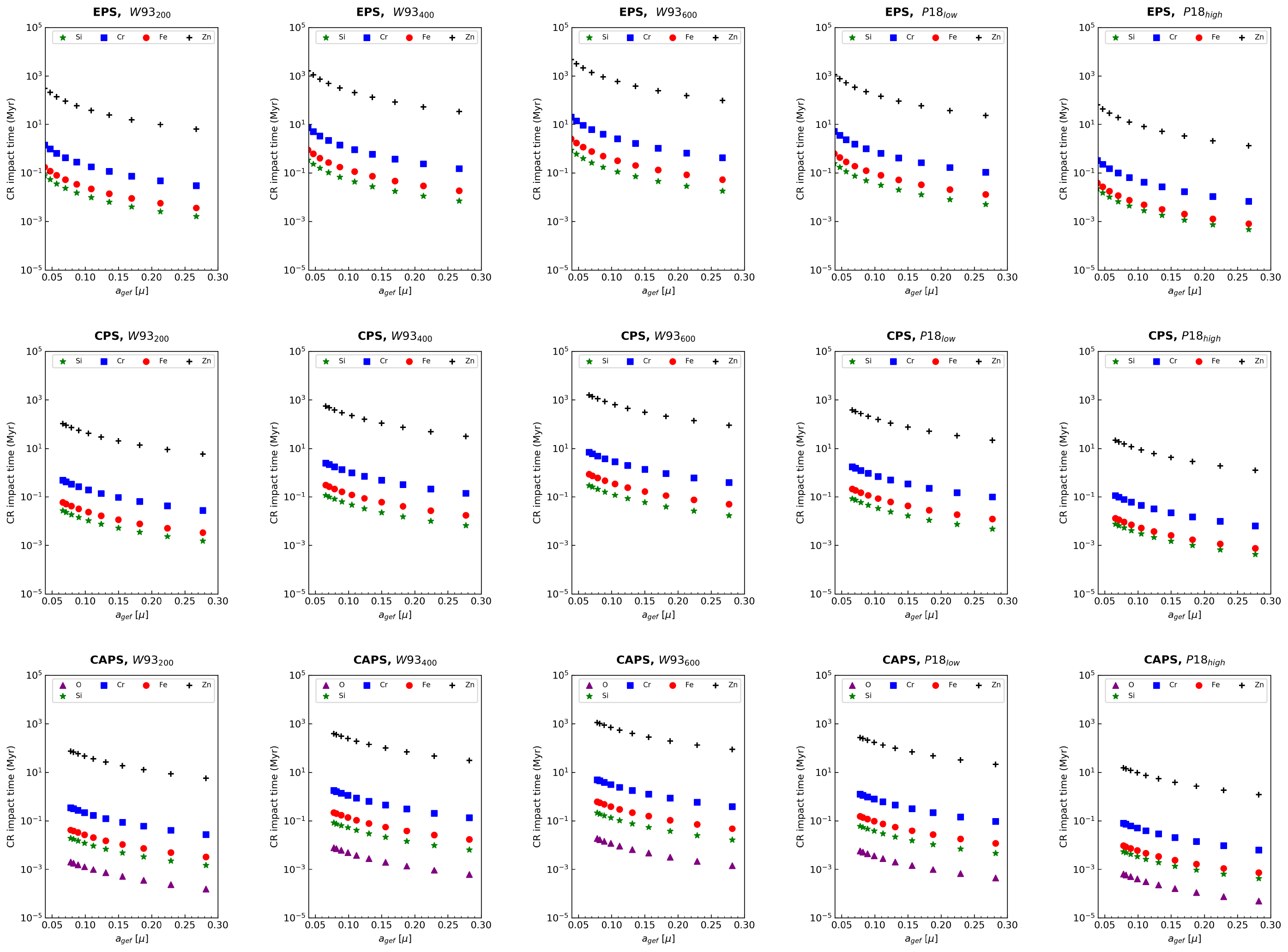}
\end{adjustbox}
\label{fig:fig10_CR_impact_time_scale}
\end{figure*}

For a specific CR ion type (i) and a grain size (k), the time between successive CR ion-grain impact (in units of s) equals  the product of the effective grain cross-section, $\rm \sigma_{\rm gef,k}$ and the CR ion flux distribution, given as follows: 
\begin{equation}\label{eq:B.2}
       \mathrm{t_{impact}} = \frac{1}{4\pi \times\sigma_{gef,k} \int_{\varepsilon_{i,min}}^{\varepsilon_{i,max}}j_{\varepsilon_i} \; d\varepsilon} \; \rm s,   
\,
   \end{equation}  
As seen in Figure \ref{fig:fig10_CR_impact_time_scale}1, the calculated impact time scales for relatively light CR ions are comparable with typical cloud core lifetime. However, when the atomic number of CR ions increases, the impact time scales tend to rise significantly because of the lower abundance of heavy CR ions.

Since larger grains have larger cross-sections for the incident CR ion, the impact time scales decrease with the increasing target grain size. In addition, the impact time scales vary notably depending on the adopted CR ion spectrum. Hence, the effect of low-energy CR ions on the energy spectrum should be included in the calculations related to CR-grain interactions.

\bsp	
\label{lastpage}
\end{document}